\documentclass{aa}
\usepackage{graphicx}
\usepackage{multirow}
\usepackage{verbatim}
\usepackage{algorithm}
\usepackage{color}

\newenvironment{ifelse}{%
  \begin{list}{}{%
      \setlength{\topsep}{0pt}\setlength{\parskip}{0pt}
      \setlength{\partopsep}{0pt}\setlength{\itemsep}{0pt}}}%
  {\end{list}}

\makeatletter

  \gdef\listctr{list\romannumeral\the\@listdepth}\expandafter
  
\makeatother

\newenvironment{AlgorithmSteps}[1][1]{%
  \begin{list}{\csname label\listctr\endcsname}{%
      \usecounter{\listctr}
      
      \settowidth{\labelwidth}{\textsc{Step\ #1.}}%
      \setlength{\leftmargin}{\labelwidth}\addtolength{\leftmargin}{\labelsep}}}%
  {\end{list}}

\begin{document}
\title{Efficient deconvolution methods for astronomical imaging:
Algorithms and IDL-GPU codes}
\author{M. Prato, R. Cavicchioli, L. Zanni\inst{1}, \\
 P. Boccacci, M. Bertero\inst{2}}
\institute{Dipartimento di Matematica Pura ed Applicata, Universit\`a
           di Modena e Reggio Emilia, Via Campi 213/b, 41125 Modena, Italy
           \and
           Dipartimento di Informatica e Scienze dell'Informazione,
           Universit\`a di Genova, Via Dodecaneso 35, 16146 Genova, Italy}
\date{Received ---; accepted ---.}

\abstract
{The Richardson-Lucy method is the most popular deconvolution
method in astronomy because it preserves the number of counts and the
non-negativity of the original object. Regularization is, in general,
obtained by an early stopping of Richardson-Lucy iterations. In the case of
point-wise objects such as binaries or open star clusters, iterations
can be pushed to convergence. However, it is well-known that Richardson-Lucy is
an inefficient method. In most cases and, in particular, for low noise
levels, acceptable solutions are obtained at the cost of hundreds or
thousands of iterations, thus several approaches to accelerating
Richardson-Lucy have been proposed. They are mainly based on Richardson-Lucy
being a scaled gradient method for the minimization of the Kullback-Leibler
divergence, or Csisz\'ar I-divergence, which represents the
data-fidelity function in the case of Poisson noise. In this framework,
a line search along the descent direction is considered for reducing
the number of iterations.}
{A general optimization method, referred to as the scaled
gradient projection method, has been proposed for the constrained
minimization of continuously differentiable convex
functions. It is applicable to the non-negative
minimization of the Kullback-Leibler divergence. If the scaling suggested by Richardson-Lucy is
used in this method, then it provides a considerable increase in the efficiency of Richardson-Lucy.
Therefore the aim of this paper is to apply the scaled
gradient projection method to a number of imaging
problems in astronomy such as single image deconvolution, multiple image
deconvolution, and boundary effect correction.}
{Deconvolution methods are proposed by applying the scaled
gradient projection method to the minimization
of the Kullback-Leibler divergence for the imaging problems mentioned above and the
corresponding algorithms are derived and implemented in interactive data language. For all the
algorithms, several stopping rules are introduced, including one based on
a recently proposed discrepancy principle for Poisson data. To attempt to achieve a
further increase in efficiency, we also consider an implementation on graphic processing
units.}
{The proposed algorithms are tested on simulated images. The acceleration of
scaled gradient projection methods achieved with respect to the corresponding Richardson-Lucy methods
strongly depends on both the problem and the specific object to be
reconstructed, and in our simulations the improvement achieved ranges from about a factor of 4 to more than 30.
Moreover, significant accelerations of up to two orders of magnitude have been observed
between the serial and parallel implementations of the algorithms.
The codes are available upon request.}
{}

\authorrunning{M. Prato et al.}
\titlerunning{Efficient deconvolution methods for astronomical imaging}
\maketitle

\section{Introduction}

The Richardson-Lucy (RL) algorithm (Richardson \cite{richardson}, Lucy
\cite{lucy}) is a renowned iterative method for image deconvolution in
astronomy and other sciences. Here, we define $\vec g$ to be the detected image and
$A$ the imaging matrix given by $A{\vec f}={\vec K}*{\vec f}$, where
${\vec K}$ is the point spread function (PSF) and $*$ denotes a convolution. 
If the PSF is then normalized to unit volume, the iteration, as modified by Snyder
(Snyder \cite{snyder}), is
\begin{equation}
{\vec f}^{(k+1)}={\vec f}^{(k)} \circ A^T \frac{{\vec g}}
{A {\vec f}^{(k)}+{\vec b}}~~,
\label{RL}
\end{equation}
where $A^T$ is the transposed matrix, ${\vec b}$ is a known array
representing background emission, ${\vec x}\circ{\vec y}$ denotes the pixel by pixel product
of two equally-sized arrays ${\vec x}$, ${\vec y}$, and
{\vec x}/{\vec y} their quotient.

It is well-known that the method has several interesting features. The
result of each iteration is non-negative and robust against small errors
in the PSF, and that flux is conserved both globally and locally if ${\vec b}=0$.

In such a case, it has also been proven by several authors
(see, for instance, Natterer \& W\"ubbeling \cite{natterer}) that the
iterations converge to either a maximum likelihood solution for Poisson data 
(Shepp \& Vardi \cite{shepp}) or, equivalently, to a
minimizer of the Kullback-Leibler (KL) divergence, which is also known as
the Csisz\'ar I-divergence (Csisz\'ar \cite{csi}), given by
\begin{eqnarray}
\label{eq:divergence}
J_0({\vec f};{\vec g})=\sum_{{\bf m} \in S}
\{{\vec g}({\bf m}) {\rm ln} \frac{{\vec g}({\bf m})}
{(A \vec{f})({\bf m})+{\vec b}({\bf m})}+\\ \nonumber
+(A \vec{f})({\bf m})+{\vec b}({\bf m})-{\vec g}({\bf m}) \}~~,
\label{KL}
\end{eqnarray}
where $S$ is the set of values of the multi-index ${\bf m}$ labeling
the image pixels.

As shown in Barrett \& Meyers (\cite{barrett}), the non-negative
minimizers of $J_0({\vec f};{\vec g})$ are sparse objects,
i.e. they consist of bright spots over a black background. Therefore,
in the case of simple astronomical objects, such as binaries
or open star clusters, the algorithm can be pushed to convergence
(examples are given in Sect. 4), while, in the case of more complex
objects, an early stopping of the iterations, providing a ``regularization
effect'', is required. The problem of introducing suitable stopping rules
is briefly discussed in Sect. 3.

The main disadvantage of the RL algorithm is that it is not very efficient:
it may require hundreds or thousands of iterations for images with a
large number of counts (low Poisson noise). In the case of large-scale
images or multiple images of the same target, the computational cost can
become prohibitive. For this reason, several acceleration schemes have
been proposed, of which we mention a few.

The first is the ``multiplicative relaxation'' proposed by Llacer \&
N\'u\~nez (\cite{llacer}), which consists in replacing the iteration
of Eq. (\ref{RL}) by
\begin{equation}
{\vec f}^{(k+1)}={\vec f}^{(k)} \circ
\left(A^T \frac{{\vec g}}{A {\vec f}^{(k)}+{\vec b}}\right)^\alpha
\label{RL1}
\end{equation}
with $\alpha > 1$. Convergence is proved in Iusem (\cite{iusem}) for
$\alpha < 2$. As demonstrated in Lant\'eri et al (\cite{lanteri0}), this
approach can provide a reduction in the number of iterations by a factor
of $\alpha$, with essentially the same cost per iteration. For low
numbers of counts numerical convergence has been found also for
$\alpha > 2$ (Anconelli et al. \cite{anco2}).
A ``linear relaxation'' is investigated in Adorf et al. (\cite{adorf}).
It can be written in the form
\begin{equation}
{\vec f}^{(k+1)}={\vec f}^{(k)} - \lambda_k{\vec f}^{(k)} \circ
\left({\vec 1} - A^T \frac{{\vec g}}{A {\vec f}^{(k)}+{\vec b}}\right)~~,
\label{RL2}
\end{equation}
where $\lambda_k >1$ (for $\lambda_k=1$ the RL algorithm is re-obtained)
and ${\vec 1}$ is the array with all entries equal to 1. Since the
quantity in brackets is the gradient of $J_0({\vec f};{\vec g})$,
we note that RL is a scaled gradient method with a
scaling given by ${\vec f}^{(k)}$ at iteration $k$, and that the
relaxation method is essentially a line search along this descent
direction, which can be performed by minimizing the objective function
$J_0({\vec f};{\vec g})$ (Adorf et al. \cite{adorf}) or applying the Armijo
rule (Lant\'eri et al. \cite{lanteri0}). A moderate
increase in efficiency is then observed by these authors. The values reached after convergence of the algorithms
can be inferred from general results of optimization theory
(Bertsekas \cite{bertsekas}).
Finally, a greater increase in efficiency on the order of ten, is observed using
an acceleration method proposed by Biggs \& Andrews (\cite{biggs}),
which exploits a suitable extrapolation along the trajectory of the
iterates, and is implemented in the {\it deconvlucy}
function of the {\it Image Processing} MATLAB toolbox. The problem with
this method is that no convergence proof is available and, in our
experience, a deviation from the trajectory of RL iterations is sometimes
observed, providing unreliable results.

Bonettini et al. (\cite{bonettini}) developed an optimization
method, which they called {\it scaled gradient projection} (SGP)
method, to constrain the minimization of a convex function, and proved that its
convergence occurs under mild conditions. This method can be quite naturally
applied to the non-negative minimization of the KL divergence, using the
scaling of the gradient suggested by RL, hence this application
of SGP can also be considered as a more efficient version of RL. In Bonettini et al.
(\cite{bonettini}), the performance of the new method is compared with
that of RL and the Biggs \& Andrews method, as implemented in MATLAB,
providing an improvement in efficiency comparable to that of the latter method, but
sometimes better and without its drawbacks. Further applications of SGP in image
restoration problems can be found e.g. in Benvenuto et al. \cite{benvenuto},
Bonettini \& Prato \cite{bonettini2010}, and Zanella et al. \cite{zanella}.

The purpose of this paper is not only to illustrate the features of SGP
to the astronomical community, but also to extend its application to the
problems of multiple image deconvolution and boundary
effect correction. The first problem is fundamental, for instance, to the
reconstruction of the images of the future interferometer of the Large
Binocular Telescope (LBT), denoted LINC-NIRVANA (Herbst et al.
\cite{herbst}), while the second problem is important in both single and
multiple image deconvolution. All the algorithms are implemented in interactive data language (IDL)
and the codes will be freely distributed. Moreover, we present an implementation for
GPU (graphic processor unit) is also provided. In this paper, we consider
only the constraint of non-negativity. Bonettini et al (\cite{bonettini}) investigated
both non-negativity and flux conservation and provided
an efficient algorithm, for computing the projection on the
convex set defined by the constraints. However, their numerical
experiments seem to demonstrate that the additional flux constraint
does not significantly improve the reconstructions.

The paper is organized as follows. In Sect. 2, after a brief description
of the general SGP algorithm in the case of non-negativity constraint, we
derive its application to the problems of both single and multiple image
deconvolution and boundary effect correction. In Sect.
3, we describe the IDL and GPU codes and in Sect. 4 we discuss our
numerical experiments illustrating the increase in efficiency achievable with the proposed
methods. In Sect. 5, we discuss possible implementation improvements and
extensions to regularized problems.

\section{The deconvolution methods}

We first describe the monotone SGP algorithm for
minimizing a convex and differentiable function on the non-negative
orthant. For the general version of the algorithm including a flux constraint, we refer
to Bonettini et al. (\cite{bonettini}). Next, we outline the application
of SGP to the three imaging problems mentioned in the Introduction.

\subsection{The scaled gradient projection (SGP) method}\label{SGPsec}

The SGP scheme is a gradient method for the solution of the problem
\begin{equation}
\min_{\vec f \geq 0} J_0({\vec f};{\vec g})~~,
\label{minpr}
\end{equation}
where $J_0({\vec f};{\vec g})$ is a convex and
continuously differentiable function defined for each one
of the problems considered in this paper. Each SGP iteration is based
on the descent direction $\vec d^{(k)} = \vec y^{(k)}-\vec f^{(k)}$,
where
\begin{equation}
{\vec y}^{(k)}=P_+(\vec f^{(k)}-\alpha_kD_k\nabla J_0({\vec f^{(k)}};{\vec g}))
\vspace{-.1cm}
\label{diry}
\end{equation}
is defined by combining a scaled steepest descent direction with
a projection on the non-negative orthant. The matrix $D_k$ in Eq. (\ref{diry})
is chosen in the set $\mathcal D$ of the $n \times n$
diagonal positive definite matrices, whose diagonal elements have values
between $L_1$ and $L_2$ for given thresholds $0 < L_1 < L_2$.

The main SGP steps are given in algorithm \ref{GPM}. The global convergence
of the algorithm is obtained by means of the standard monotone Armijo rule
in the line--search procedure described in step 5 (see Bonettini et al.
\cite{bonettini}).

\begin{algorithm}[h]
\caption{Scaled gradient projection (SGP) method}
\label{GPM}
Choose the starting point $\vec f^{(0)} \geq 0$ and set the parameters $\beta, \theta\in (0,1)$,
$0< \alpha_{min} <\alpha_{max}$.\\[.2cm]
{\textsc{For}} $k=0,1,2,...$ \textsc{do the following steps:}
\begin{itemize}
\item[]
\begin{AlgorithmSteps}[4]
\item[1] Choose the parameter $\alpha_k \in [\alpha_{min},\alpha_{max}]$ and the scaling matrix $D_k\in \mathcal D$;
\item[2] Projection: $$\vec y^{(k)} = P_+(\vec f^{(k)}-\alpha_kD_k\nabla J_0({\vec f^{(k)}};{\vec g}));$$
\item[3] Descent direction: ${\vec d^{(k)}} = {\vec y^{(k)}} - {\vec f^{(k)}}$;
\item[4] Set $\lambda_k = 1$;
\item[5] Backtracking loop:
\begin{ifelse}
\item let $J_{new} = J_0(\vec f^{(k)} + \lambda_k \vec d^{(k)};{\vec g})$;
\item \textsc{If} $$J_{new}\leq
           J_0({\vec f^{(k)}};{\vec g})+\beta\lambda_k\nabla J_0({\vec f^{(k)}};{\vec g})^T \vec d^{(k)}$$
      \textsc{then} \\  \hspace*{.5cm} go to step 6;
\item \textsc{Else} \\  \hspace*{.5cm} set $\lambda_k = \theta \lambda_k$ and go to step 5.
\item \textsc{Endif}
\end{ifelse}
\item[6] Set $\vec f^{(k+1)} = \vec f^{(k)} + \lambda_k \vec d^{(k)}$.
\end{AlgorithmSteps}
\end{itemize}
\textsc{End}
\end{algorithm}

We emphasize that any choice of the
steplength $\alpha_k \in [\alpha_{min},\alpha_{max}]$ and the scaling
matrix $D_k \in \mathcal D$ are allowed; this freedom of choice can then
be fruitfully exploited for introducing performance improvements. \\
An effective selection strategy for the steplength parameter is obtained
by adapting to the context of the scaling gradient methods the Barzilai and
Borwein (\cite{barzilai}) rules (hereafter denoted BB), which are
widely used in standard nonscaled gradient methods. When the scaled
direction $D_k\nabla J_0({\vec f^{(k)}};{\vec g})$ is exploited within a
step of the form
$(\vec f^{(k)}-\alpha_kD_k\nabla J_0({\vec f^{(k)}};{\vec g}))$, the BB
steplength rules become
\begin{equation}
\alpha_k^{{(BB1)}} = \frac{{\vec{s}^{(k-1)}}^T D_k^{-1} D_k^{-1}
\vec{s}^{(k-1)}}{{\vec{s}^{(k-1)}}^T D_k^{-1} \vec{z}^{(k-1)}}\,,
\label{BB1}
\end{equation}
\begin{equation}
\alpha_k^{{(BB2)}} = \frac{{\vec{s}^{(k-1)}}^T D_k \vec{z}^{(k-1)} }
{{\vec{z}^{(k-1)}}^T D_k D_k \vec{z}^{(k-1)}}\,,
\label{BB2}
\end{equation}
where $\vec{s}^{(k-1)}\!\!=\!\!\vec{f}^{(k)}\!\! -\!\! \vec{f}^{(k-1)}$ and
$\vec{z}^{(k-1)}\!\!=\!\!\nabla J_0({\vec f^{(k)}};{\vec g})\!\!-\!\!\nabla J_0({\vec f^{(k-1)}};{\vec g})$.
In SGP, we constrain the values produced by these rules into the interval $[\alpha_{min},\alpha_{max}]$ in the following way:
\vskip0.2cm
\noindent
\hspace*{0.3cm}
\textsc{if} $ {\vec{s}^{(k-1)}}^T D_k^{-1} \vec{z}^{(k-1)} \le 0$ \textsc{then}\\
\hspace*{.5cm} $\alpha_k^{(1)} = \min\left\{10 \cdot \alpha_{k-1}, \ \alpha_{max} \right\}$; \\
\hspace*{.3cm}
\textsc{else}\\
\hspace*{.5cm} $\alpha_k^{(1)} = \min\left\{\alpha_{max}, \ \max\left\{ \alpha_{min}, \ \alpha_k^{{(BB1)}} \right\}\right\}$; \\
\hspace*{.3cm}
\textsc{endif} \\
\hspace*{0.3cm}
\textsc{if} $ {\vec{s}^{(k-1)}}^T D_k \vec{z}^{(k-1)} \le 0$ \textsc{then}\\
\hspace*{.5cm} $\alpha_k^{(2)} = \min\left\{10 \cdot \alpha_{k-1}, \ \alpha_{max} \right\}$; \\
\hspace*{.3cm}
\textsc{else}\\
\hspace*{.5cm} $\alpha_k^{(2)} = \min\left\{\alpha_{max}, \ \max\left\{ \alpha_{min}, \ \alpha_k^{{(BB2)}} \right\}\right\}$; \\
\hspace*{.3cm}
\textsc{endif}
\\[.1cm]
The recent literature on steplength selection in gradient methods propose that 
steplength updating rules be designed by alternating the two BB formulae (Serafini et al. \cite{serafini}, Zhou et al. \cite{zhou}).
In the case of nonscaled gradient methods (i.e., $D_k=I$) where the inequality $\alpha_k^{(BB2)} \le \alpha_k^{(BB1)}$ holds (Serafini et al. \cite{serafini}),
remarkable convergence rate improvements have been obtained by alternation strategies that force the selection to be made in a suitable order of both low and high BB values. In Frassoldati et al. (\cite{frassoldati}), this aim is realized by an
alternation criterion, which compares well with other popular BB-like steplength rules, namely
\vskip0.2cm
\noindent
\hspace*{0.3cm}
\textsc{if} $ {\alpha_k^{(2)}}/{\alpha_k^{(1)}} \le
\tau_k$ \textsc{then}
\begin{equation} \hspace*{.6cm} \alpha_k = \min_{j=\max\left\{1,k+1-M_{\alpha}\right\},\dots,k} \ \alpha_j^{(2)};\label{alphadef}\end{equation}
\hspace*{.5cm} $\tau_{k+1} = 0.9 \cdot \tau_{k}$;\\
\hspace*{.3cm}
\textsc{else}\\
\hspace*{.5cm} $\alpha_k = \alpha_k^{(1)}; \qquad \tau_{k+1} = 1.1 \cdot \tau_{k}$;\\
\hspace*{.3cm}
\textsc{endif}
\\[.1cm]
where $M_{\alpha}$ is a prefixed positive integer and $\tau_1\in (0,1)$.
When scaled versions of the BB rules given in Eqs. (\ref{BB1})-(\ref{BB2}) are used, the inequality $\alpha_k^{(BB2)} \le \alpha_k^{(BB1)}$ is not always true. Nevertheless, a wide computational study suggests that this alternation criterion is more suitable in terms of
convergence rate than the use of a single BB rule (Bonettini et al. \cite{bonettini}, Favati et al. \cite{favati}, Zanella et al. \cite{zanella}).
Furthermore, in our experience, the use of the BB values provided by Eq. (\ref{alphadef})
in the first iterations slightly improves the reconstruction accuracy and, consequently,
in the proposed SGP version we start the steplength alternation only after the first 20 iterations.

When selecting the scaling matrix $D_k$, a suitable updating rule
generally depends on the special form of the objective function. In our case,
we chose the scaling matrix suggested by the RL algorithm, i.e.,
\begin{equation}
D_k = {\rm diag}\left(\min\left[L_2, \ \max\left\{L_1,
{\vec f^{(k)}}\right\} \right]\right)~,
\label{Dk}
\end{equation}
where $L_1,L_2$ are prefixed thresholds.

\subsection{Single image deconvolution}

The problem of single image deconvolution in the presence of photon counting noise is
the minimization of the KL divergence defined in Eq. (\ref{eq:divergence})
and the solution is given by the iterative RL algorithm of Eq. (\ref{RL}).
When applying SGP, we only need the expression of the gradient of the KL
divergence, which is given by (when the normalization of the
PSF to unit volume is used)
\begin{equation}
\nabla J_0({\vec f};{\vec g})= {\vec 1}-A^T \frac{{\vec g}}
{A{\vec f}+ {\vec b}}~~.
\label{KLgrad}
\end{equation}
The SGP behavior with respect to RL was previously investigated in Bonettini et al.
(\cite{bonettini}).

\subsection{Multiple image deconvolution}

Successful multiple image deconvolution is fundamental to the future
Fizeau interferometer of LBT called LINC-NIRVANA (Herbst et al.
\cite{herbst}) or to the ``co-adding'' method of images with
different PSFs proposed by Lucy \& Hook (\cite{lucyhook}).

We define $p$ to be the number of detected images ${\vec g}_j$, ($j$=1,..,$p$),
with corresponding PSFs ${\vec K}_j$, all normalized to unit volume,
and $A_j{\vec f}= {\vec K}_j*{\vec f}$. It is quite natural
to assume that the $p$ images are statistically independent, such that
the likelihood of the problem is the product of the likelihoods of the
different images. If we assume again Poisson statistics, and we take the
negative logarithm of the likelihood, then the maximization of the
likelihood is equivalent to the minimization of a data-fidelity function,
which is the sum of KL divergences, one for each image, i.e.
\begin{eqnarray}
\label{multipleKL}
J_0({\vec f};{\vec g})=\sum_{j=1}^{p}\sum_{{\bf m} \in S}
\{{\vec g}_j({\bf m}) {\rm ln} \frac{{\vec g}_j({\bf m})}
{(A_j \vec{f})({\bf m})+{\vec b}_j({\bf m})}+\\ \nonumber
+(A_j \vec{f})({\bf m})+{\vec b}_j({\bf m})-{\vec g}_j({\bf m}) \}~~.
\end{eqnarray}
If we apply the standard expectation maximization method (Shepp \&
Vardi \cite{shepp}) to this problem, we obtain the iterative
algorithm
\begin{equation}
{\vec f}^{(k+1)}=\frac{1}{p}{\vec f}^{(k)} \circ \sum_{j=1}^p
A_j^T \frac{{\vec g}_j}{A_j{\vec f}^{(k)}+{\vec b}_j}~~,
\label{multipleRL}
\end{equation}
which we call the {\it multiple image} RL method (multiple RL, for short).
Since the gradient of (\ref{multipleKL}) is given by
\begin{equation}
\nabla J_0({\vec f};{\vec g})=\sum_{j=1}^{p}\left\{{\vec 1} -
A_j^T \frac{{\vec g}_j}{A_j{\vec f}+{\vec b}_j} \right\}~~,
\label{multiplegradient}
\end{equation}
we find that the algorithm presented in Eq. (\ref{multipleRL}) is a scaled gradient
method, with a scaling given, at iteration $k$, by ${\vec f}^{(k)}/p$.
Therefore, the application of SGP to this problem is straightforward.

However, for the reconstruction of LINC-NIRVANA images, we must consider
that an acceleration of the algorithm in Eq. (\ref{multipleRL}) is proposed in
Bertero \& Boccacci (\cite{bertero2000}) by exploiting an analogy
between the images of the interferometer and the projections in
tomography. In this approach called OSEM (ordered subset expectation
maximization, Hudson \& Larkin \cite{hudson}),
the sum over the $p$ images in Eq. (\ref{multipleRL}) is replaced by a
cycle over the same images. To avoid oscillations of the
reconstructions within the cycle, a preliminary step is the
normalization of the different images to the same flux, if different
integration times are used in the acquisition process.
The method OSEM is summarized in algorithm \ref{OSEM}.

\begin{algorithm}[t]
\caption{Ordered subset expectation
maximization (OSEM) method}
\label{OSEM}
Choose the starting point $\vec f^{(0)} > 0$.\\[.2cm]
{\textsc{For}} $k=0,1,2,...$ \textsc{do the following steps:}
\begin{itemize}
\item[]
\begin{AlgorithmSteps}[4]
\item[1] Set ${\vec h}^{(0)}={\vec{f}}^{(k)}$;
\item[2] {\textsc{For}} $j=1,...,p$ \textsc{compute}
\begin{equation}
{\vec h}^{(j)}={\vec h}^{(j-1)}\circ
\left({A}_j^T \frac{{\vec g}_j}{{A}_j {\vec h}^{(j-1)}+
{\vec b}_j} \right)~~;
\label{OSEMstep}
\end{equation}
\item[3] Set ${\vec{f}}^{(k+1)}={\vec h}^{(p)}$.
\end{AlgorithmSteps}
\end{itemize}
\textsc{End}
\end{algorithm}
As follows from practice and theoretical remarks, this approach
reduces the number of iterations by a factor $p$. However, the
computational cost of one multiple RL iteration is lower than that of
one OSEM iteration: we need $3p+1$ FFTs in the first case and $4p$ FFTs
in the second. In conclusion, the increase in efficiency provided by OSEM is
roughly given by $(3p+1)/4$. When $p=3$ (the number of images provided
by the interferometer will presumably be small), the efficiency is higher by a factor of 2.5,
and a factor of 4.7 when $p=6$. These results must be taken into account
when considering the increase in the efficiency of SGP with respect to multiple RL. We
can add that the convergence of SGP is proven while that of OSEM is not,
even if it has always been verified in our numerical experiments.

\subsection{Boundary effect correction}\label{bound}

If the target ${\vec f}$ is not completely contained in the image domain,
then the previous deconvolution methods produce annoying boundary
artifacts. It is not the purpose of this paper to discuss the different
methods for solving this problem. We focus on an approach proposed in
Bertero \& Boccacci (\cite{bertero2005}) for single image deconvolution
and in Anconelli et al. (\cite{anco3}) for multiple image deconvolution.
Here we present the equations in the case of multiple images, where a single image
corresponds to $p=1$.

The idea is to reconstruct the object ${\vec f}$ over a domain broader
than that of the detected images and to merge, by zero padding, the
arrays of the images and the object into arrays of dimensions
that enable their Fourier transform to be computed by means of FFT. We denote 
by $\bar S$ the set of values of the multi-index labeling the pixels of
the broader arrays containing $S$ and by $R$ that of the object array contributing to $S$,
such that $S \subset R \subset \bar S$. It is also obvious
that also the PSFs must be defined over $\bar S$ and that this can be done in
different ways, depending on the specific problem one is considering.
We point out that they must be normalized to unit volume over $\bar S$.
We also note that $R$ corresponds to the part of the object contributing to the
detected images and that it depends on the extent of the PSFs. It can be
estimated from this information as we indicate in the following (see Eq. (\ref{Rdef})).
The reconstruction of ${\vec f}$ outside $S$, is unreliable in most
cases, but its reconstruction inside $S$ is practically free of boundary
artifacts, as shown in the papers cited above and in the experiments of
Sect. 4.

If we denote by ${\vec M}_R$, ${\vec M}_S$ the arrays, defined over
$\bar S$, which are 1 over $R$, $S$ respectively and 0 outside, we define
the matrices $A_j$ and $A_j^T$
\begin{eqnarray}
\label{projection}
(A_j \vec{f})({\bf m})={\vec M}_S({\bf m})
\sum_{{\bf n} \in \bar S}{\vec K}_j({\bf{m-n}})
{\vec M}_R({\bf n}) {\vec f}({\bf n})~~, \\
\label{backprojection}
(A_j^T \vec{g})({\bf n})={\vec M}_R({\bf n})
\sum_{{\bf m} \in \bar S}{\vec K}_j({\bf{m-n}}){\vec M}_S ({\bf m}){\vec g}
({\bf m})~.
\end{eqnarray}
In the second equation, $\vec g$ denotes a generic array defined over
$\bar S$. Both matrices can be easily computed by means of FFT.
With these definitions, the data fidelity function is then given
again by Eq. (\ref{multipleKL}), with $S$ replaced by $\bar S$, while
its gradient is now given by
\begin{equation}
\label{gradientboundary}
\nabla J_0({\vec f};{\vec g})=\sum_{j=1}^{p}\left\{ A_j^T {\vec 1} -
A_j^T \frac{{\vec g}_j}{A_j{\vec f}+{\vec b}_j} \right\}~~,
\end{equation}
leading to the introduction of the functions
\begin{eqnarray}
& &{\vec \alpha}_j({\bf n})=\sum_{{\bf m} \in \bar S}{\vec K}_j({\bf{m-n}}){\vec M}_S ({\bf m}){\vec g}
({\bf m})~~, \\
\nonumber
& &{\vec \alpha}({\bf n})=\sum_{j=1}^p{\vec \alpha}_j({\bf n})~~,~~{\vec n} \in \bar{S}~~.
\end{eqnarray}
These functions can be used to define the
reconstruction domain $R$, since they can be either very small or zero in
pixels of $\bar S$, depending on the behavior of the PSFs. Given a
thresholding value $\sigma$, we use the definition
\begin{equation}
R=\{{\vec n} \in {\bar S}~|~ {\vec \alpha}_j({\vec n}) \geq \sigma; j=1,..,p\}~~.
\label{Rdef}
\end{equation}
The RL algorithm, with boundary effect correction, is then given by
\begin{equation}
\label{boundaryRL}
{\vec f}^{(k+1)}=\frac{{\vec M}_R}{{\vec \alpha}} \circ {\vec f}^{(k)}
\circ \sum_{j=1}^p A_j^T \frac{{\vec g}_j}{A_j{\vec f}^{(k)}+{\vec b}_j}~~,
\end{equation}
the quotient being zero in the pixels outside $R$.
Similarly, the OSEM algorithm, with a boundary effect correction is given
by algorithm \ref{OSEM} where Eq. (\ref{OSEMstep}) is replaced by
\begin{equation}
\label{boundaryOSEM}
{\vec h}^{(j)}=\frac{{\vec M}_R}{{\vec \alpha}_j} \circ {\vec h}^{(j-1)}
\circ \left({A}_j^T \frac{{\vec g}_j}{{A}_j {\vec h}^{(j-1)}+
{\vec b}_j} \right)~~.
\end{equation}
As far as the SGP algorithm concerns, the boundary effect correction is
incorporated to the scaling matrix
\begin{equation}
D_k = {\rm diag}\left(\frac{{\vec M}_R}{{\vec \alpha}} \circ
\min\left[L_2, \ \max\left\{L_1,{\vec f^{(k)}}\right\} \right]\right),
\label{Dkbound}
\end{equation}
while all the other steps remain unchanged.

\section{Computational features}

The description of the SGP algorithm provided in Sec. \ref{SGPsec} indicates several ingredients on which the success of the recipe
depends: the choice of the starting point, the selection of the
parameters defining the method, and the stopping criterion. In the following,
we briefly describe which choices were made in our numerical
experimentation, and comment on the parallel implementation
of the algorithm.

\subsection{Initialization}

As far as the SGP initial point ${\vec f}^{(0)}$ concerns, any non-negative image
is allowed. The possible choices implemented in our code are:
\begin{itemize}
\item the null image ${\vec f}^{(0)} = {\vec 0}$.
\item the noisy image ${\vec g}$ (or, in the case of multiple deconvolution,
the noisy image ${\vec g}_1$ corresponding to the first PSF ${\vec K}_1$).
\item a constant image with pixel values equal to the background-subtracted
flux (or mean flux in the case of multiple deconvolution) of the noisy data
divided by the number of pixels. If the boundary effect correction is
considered, only the pixels in the object array $R$ become equal to this
constant, while the remaining values of ${\bar S}$ are set to zero.
In future extensions of our codes, the constant image will
be convolved with a Gaussian to avoid the presence of sharp edges.
\item any input image provided by the user.
\end{itemize}
The constant image ${\vec f}^{(0)}$ was chosen for our numerical
experiments, which is also the initial point used for RL.

\subsection{SGP parameter setting}

Even if the number of SGP parameters is certainly higher than those of the RL
and OSEM approaches, the huge amount of tests carried out in several
applications has led to an optimization of these values, which allows the user
to have at his disposal a robust approach without the need for any
problem-dependent parameter tuning. In our present study, some of these values were fixed
according to the original paper of Bonettini et al. (\cite{bonettini}),
as in the case of the line-search parameters $\beta$ and $\theta$, which were set
to $10^{-4}$ and $0.4$, respectively. In addition, most of the steplength parameters
remained unchanged, as $\alpha_0 = 1.3$, $\tau_1 = 0.5$, $\alpha_{max}=10^5$,
and $M_\alpha=3$, while $\alpha_{min}$ was set to $10^{-5}$.

The main change concerned the choice of the bounds $(L_1,L_2)$ for the
scaling matrices. While in the original paper, the choice
was a couple of fixed values $(10^{-10},10^{10})$, independent of the
data, we decided to automatically adapt these bounds to the input image:
we performed one step of the RL method and tuned the parameters $(L_1,L_2)$
according to the min/max positive values $y_{\min}$/$y_{\max}$
of the resulting image according to the rule
\vskip0.2cm
\noindent
\hspace*{0.3cm}
\textsc{if} $ y_{\max}/y_{\min} < 50$ \textsc{then}\\
\hspace*{.5cm} $L_1 = y_{\min}/10$;\\
\hspace*{.5cm} $L_2 = y_{\max} \cdot 10$;\\
\hspace*{.3cm}
\textsc{else}\\
\hspace*{.5cm} $L_1 = y_{\min}$;\\
\hspace*{.5cm} $L_2 = y_{\max}$;\\
\hspace*{.3cm}
\textsc{endif}

\subsection{Stopping rules}

As mentioned in the introduction, in many instances both
RL and SGP must not be pushed to convergence and an early stopping of the
iterations is required to achieve reasonable reconstructions. In our
code, we introduced different stopping criteria, which can be adapted
by the user according to his/her purposes:
\begin{itemize}
\item fixed number of iterations. The user can decide how many iterations
of SGP must be done.
\item convergence of the algorithm. In such a case, a stopping criterion
based on the convergence of the data-fidelity function to its minimum value
is introduced. Iteration is stopped when
\begin{equation}
|J_0({\vec f}^{(k+1)};{\vec g})-J_0({\vec f}^{(k)};{\vec g})|
\leq tol~J_0({\vec f}^{(k)};{\vec g})~~,
\label{convergence}
\end{equation}
where $tol$ is a parameter that can be selected by the user.
\item minimization of the reconstruction error. This criterion can be used
in a simulation study. If one knows the object $\widetilde{\vec f}$ used to
generate the synthetic images, then one can stop the iterations when the
relative reconstruction error
\begin{equation}
\rho^{(k)}=\frac{|{\vec f}^{(k)}-\widetilde{\vec f}|}{|\widetilde{\vec f}|}
\end{equation}
reaches a minimum value. A very frequently used measure of error is
given by the $\ell_2$ norm, i.e. $|\!\cdot\!|=\|\!\cdot\!\|_2$ and this is the
criterion implemented in our code.
\item the use of a discrepancy criterion. In the case of real data, one can use a
given value of some measure of the ``distance'' between the real data and
the data computed by means of the current iteration. A recently proposed
criterion consists in defining the ``discrepancy function'' for
$p$ images ${\vec g}_j$ of size $N \times N$
\begin{equation}
\mathcal{D}^{(k)}=\frac{2}{p~N^2}J_0({\vec f}^{(k)};{\vec g})~~,
\label{discrepancy}
\end{equation}
and stopping the iterations when $\mathcal{D}^{(k)}=b$, where $b$ is a given
number close to 1. Work is in progress to validate this discrepancy 
criterion with the purpose of obtaining rules of thumb for estimating $b$ in real applications.

\end{itemize}
The last stopping rule deserves a few comments. In Bertero et al.
(\cite{bertero2010}), it is shown that, for a single image,
if $\widetilde{\vec f}$ is the object generating the noisy image ${\vec g}$,
then the expected value of $J_0(\widetilde{\vec f};{\vec g})$ is close
to $N^2/2$. This property is used to select a value of the
regularization parameter when the image reconstruction problem is
formulated as the minimization of the KL divergence with the addition
of a suitable regularization term.
This use is justified by evidence that in some important cases it provides a
unique value of the regularization parameter. Moreover, it has also been
shown that the quantity $\mathcal{D}^{(k)}$, defined in Eq. (\ref{discrepancy}),
decreases for increasing $k$, starting from a value greater than 1.
Therefore, it can be used as a stopping criterion. Preliminary
numerical experiments described in that paper show that it can provide
a sensible stopping rule at least in simulation studies.

\subsection{IDL and GPU implementation}

\newcommand{\Cpp}{\mbox{C\raisebox{0.1em}{\scriptsize ++}}}
\newcommand{\CUDA}{\mbox{CUDA}}

Our implementation of the SGP algorithm was written in IDL, a well-known and frequently
used language in astronomical environments. This
data-analysis programming language is well-suited to work with images,
using optimized built-in vector operations. Nevertheless, it is not intended that usability should be compromised by 
performance.

As already shown in Ruggiero et al (\cite{ruggiero}), the \Cpp\
implementation of the SGP algorithm
is well-suited to parallelization and good computational speedup is obtained
exploiting the \CUDA\ technology. The \CUDA\ is a framework developed by NVIDIA that enables the use of graphic processing unit
(GPU) for programming. These graphics cards are nowadays
in many personal computers and their core is highly parallel, consisting of several
hundreds of computational units. Many recent applications show that the
increase in efficiency achieved with this technology is significant and its cost is much
lower than that of a medium-sized cluster. We note that memory management is
crucial to ensure the optimal
performance when using GPU. The transfer speed of data from central memory
to GPU is much slower than the GPU-to-GPU transfer, hence to maximize the
GPU benefits it is very important to reduce the CPU-to-GPU memory
communications and retain all the problem data in the GPU memory.

The \CUDA\ technology is available in IDL as part of GPUlib, a software library
that enables GPU-accelerated calculations, developed by Tech-X Corporation.
It has to be noted that the FFT routine included in the current
version of GPUlib (1.4.4) is available only in single precision. Results from this
function differ slightly from the ones obtained in double precision
by IDL, causing some numerical differences in our experiments.

\section{Results}

\begin{table*}
\caption{Iteration numbers, relative r.m.s. errors,
computational times, and speedups of RL and SGP, provided by the corresponding
GPU implementations, for the three $256 \times 256$ objects nebula, galaxy
and Crab. Iterations are stopped at a minimum relative r.m.s. error in the serial
algorithms (the asterisks denote the maximum number of iterations allowed)}.
\label{t1}
\centering
\begin{tabular}{c c c c c c c c c c c c c}
\hline\hline
          & \multicolumn{4}{c}{Nebula ($m=10$)}  & \multicolumn{4}{c}{Galaxy ($m=10$)}   & \multicolumn{4}{c}{Crab ($m=10$)} \\
Algorithm & It  & Err   & Sec   & SpUp       & It        & Err   & Sec   & SpUp  & It   & Err   & Sec   & SpUp   \\
\hline
RL 			  & 528 & 0.021 & 41.28 &    -        & 10000$^*$ & 0.140 & 795.3 &    -   & 5353 & 0.128 & 419.8 &    -    \\
RL\_CUDA  & 528 & 0.021 & 2.079 & 19.9        & 10000$^*$ & 0.140 & 35.09 & 22.7   & 5353 & 0.128 & 19.45 & 21.6    \\
SGP 		  &  50 & 0.021 & 4.719 &    -        &       406 & 0.141 & 38.61 &    -   &  151 & 0.129 & 14.28 &    -    \\
SGP\_CUDA &  50 & 0.021 & 0.344 & 13.7        &       406 & 0.142 & 3.313 & 11.7   &  151 & 0.129 & 1.219 & 11.7    \\
\hline
          & \multicolumn{4}{c}{Nebula ($m=12$)}  & \multicolumn{4}{c}{Galaxy ($m=12$)}   & \multicolumn{4}{c}{Crab ($m=12$)} \\
Algorithm & It  & Err   & Sec   & SpUp       & It        & Err   & Sec   & SpUp  & It   & Err   & Sec   & SpUp \\
\hline
RL 			  & 124 & 0.026 & 9.797 &    -        &      3887 & 0.157 & 304.6 &    -   &  954 & 0.136 & 74.83 &    -    \\
RL\_CUDA  & 124 & 0.026 & 0.516 & 19.0        &      3887 & 0.157 & 14.50 & 21.0   &  954 & 0.136 & 3.516 & 21.3    \\
SGP 		  &  24 & 0.026 & 2.344 &    -        &       153 & 0.159 & 14.42 &    -   &   52 & 0.137 & 4.984 &    -    \\
SGP\_CUDA &  24 & 0.026 & 0.203 & 11.5        &       153 & 0.159 & 1.266 & 11.4   &   52 & 0.137 & 0.406 & 12.3    \\
\hline
          & \multicolumn{4}{c}{Nebula ($m=15$)}  & \multicolumn{4}{c}{Galaxy ($m=15$)}   & \multicolumn{4}{c}{Crab ($m=15$)} \\
Algorithm & It  & Err   & Sec   & SpUp       & It        & Err   & Sec   & SpUp  & It   & Err   & Sec   & SpUp   \\
\hline
RL 			  & 124 & 0.063 & 9.766 &    -        &       448 & 0.234 & 35.14 &    -   &  128 & 0.172 & 10.09 &    -    \\
RL\_CUDA  & 124 & 0.063 & 0.469 & 20.8        &       448 & 0.234 & 1.594 & 22.0   &  128 & 0.172 & 0.483 & 20.9    \\
SGP 		  &  12 & 0.060 & 1.250 &    -        &        21 & 0.234 & 2.094 &    -   &   10 & 0.172 & 1.093 &    -    \\
SGP\_CUDA &  12 & 0.060 & 0.109 & 11.5        &        21 & 0.234 & 0.156 & 13.4   &   10 & 0.172 & 0.093 & 11.8    \\
\hline
\end{tabular}
\end{table*}

We now demonstrate, by means of a few numerical experiments, the
effectiveness of the SGP algorithm and its IDL--based GPU implementation in
the solution of the deblurring problems described in Sect. 2. Our test
platform consists in a personal computer equipped with an AMD Athlon X2
Dual-Core at 3.11GHz, 3GB of RAM, and the graphics processing unit NVIDIA GTX
280 with \CUDA\ 3.2. We consider CPU implementations of RL, SGP, and OSEM in
IDL 7.0; the GPU implementations are developed in mixed IDL and \CUDA\
language by means of the GPUlib 1.4.4.

The set of numerical experiments can be divided into two groups: single image
and multiple image deconvolution. For each group, some tests on boundary
effect correction are included.

\subsection{Single image deconvolution}\label{s41}

\begin{figure}
	\centering
		\begin{tabular}{cc}
			\includegraphics[scale=0.3]{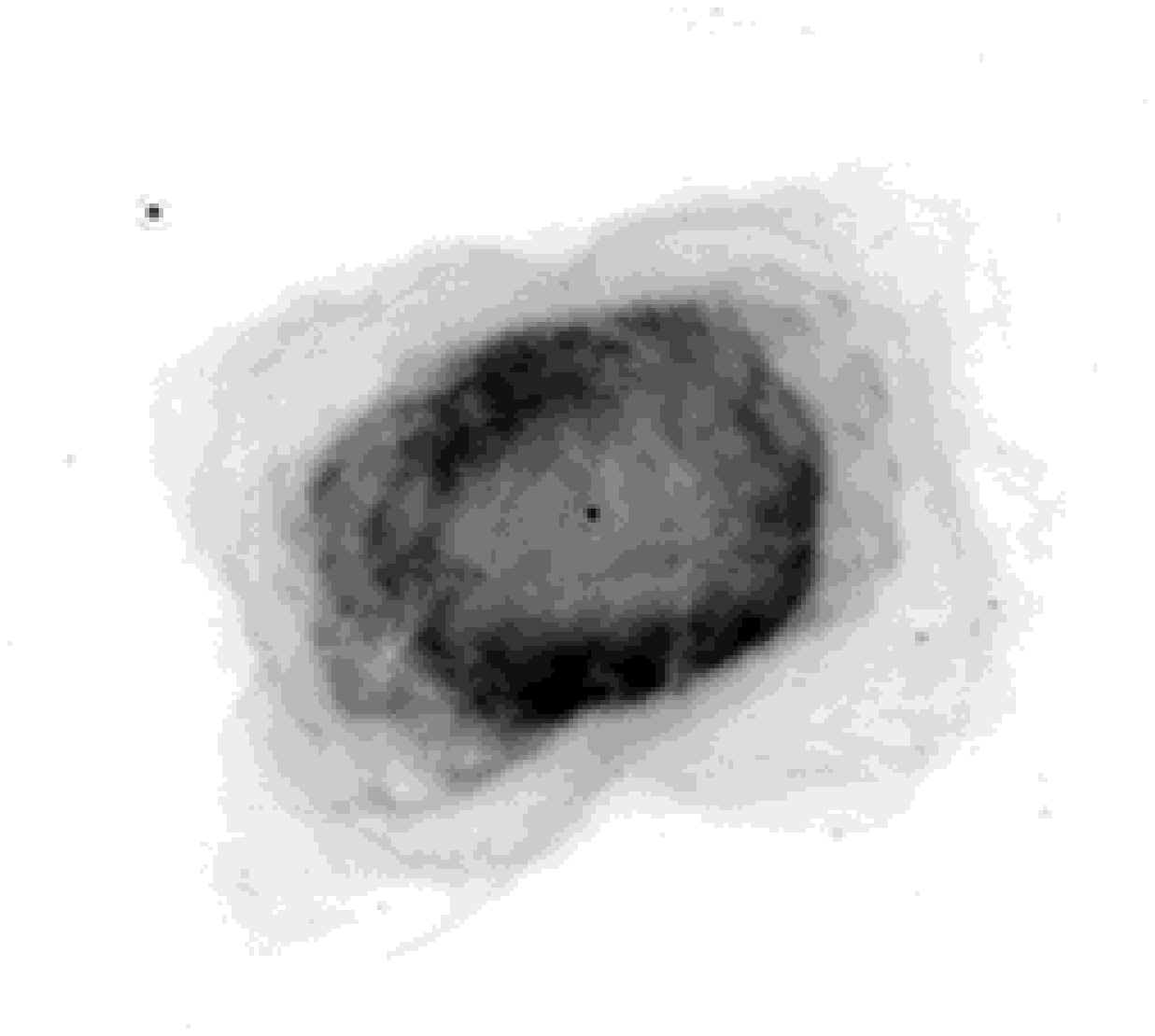} &
		  \includegraphics[scale=0.3]{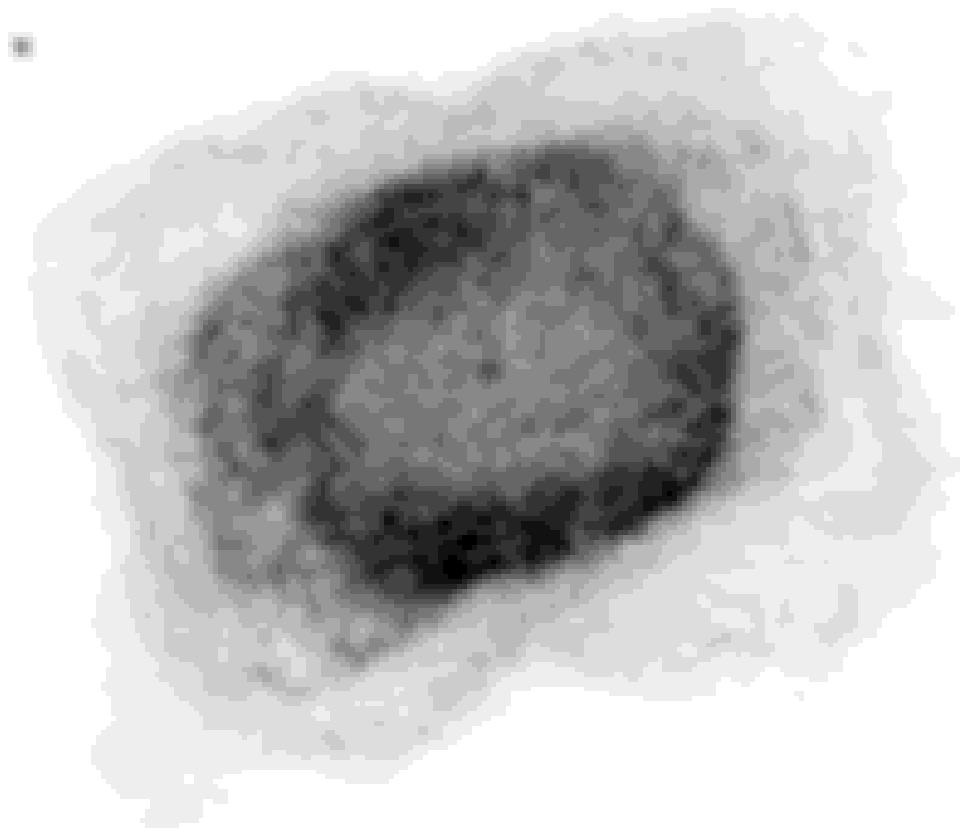} \\
			\includegraphics[scale=0.3]{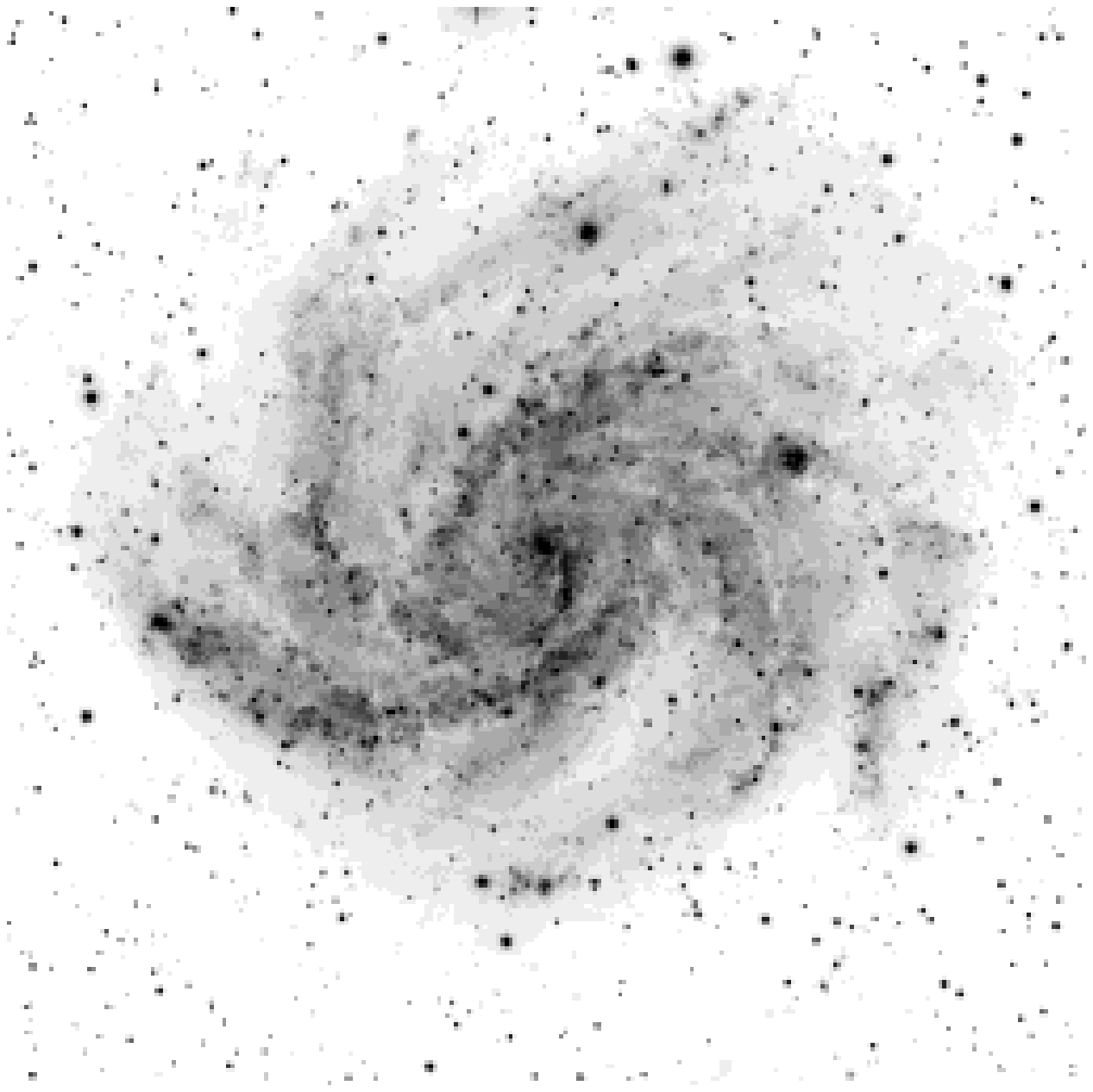} &
		  \includegraphics[scale=0.3]{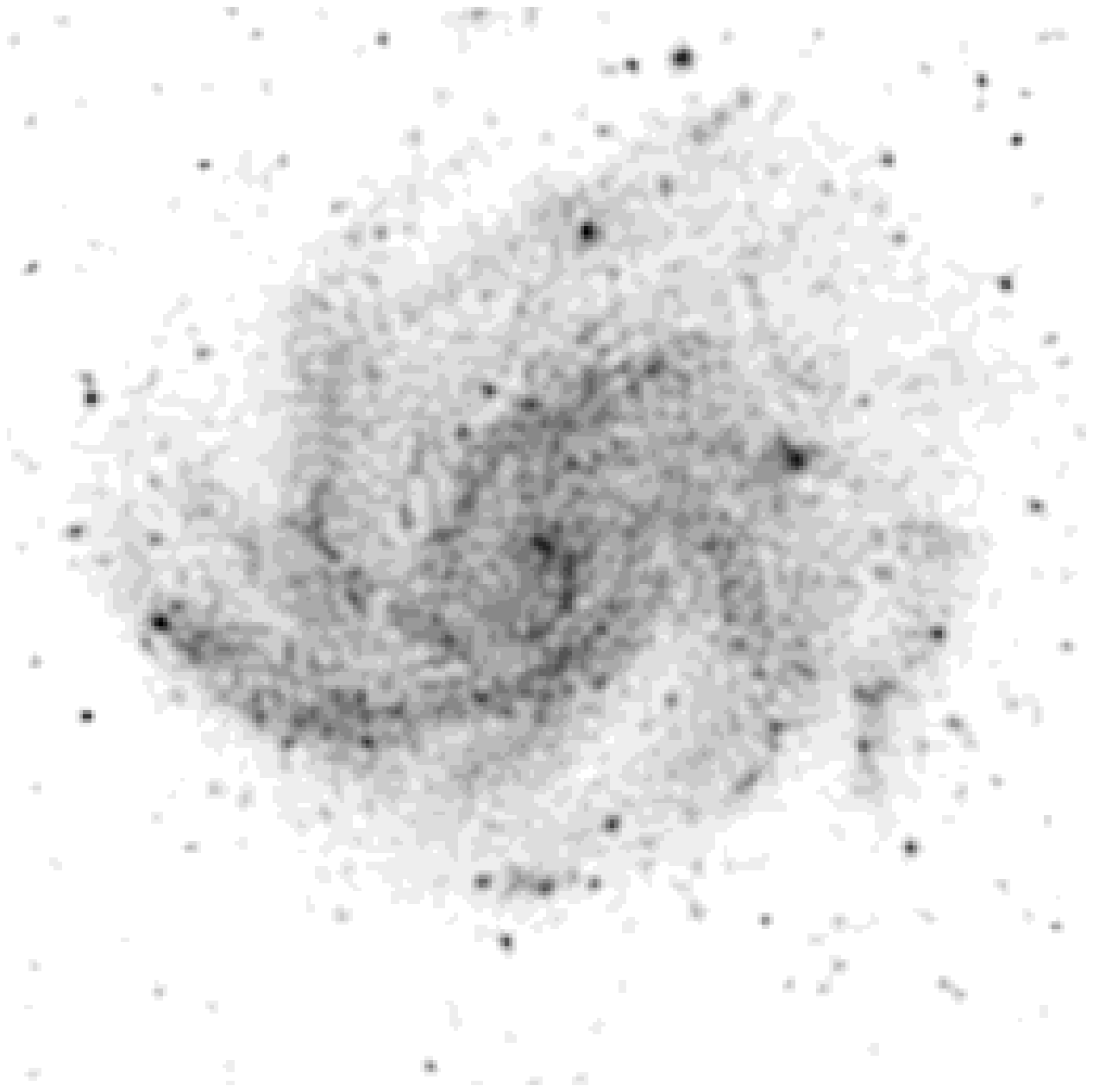} \\
		  \includegraphics[scale=0.3]{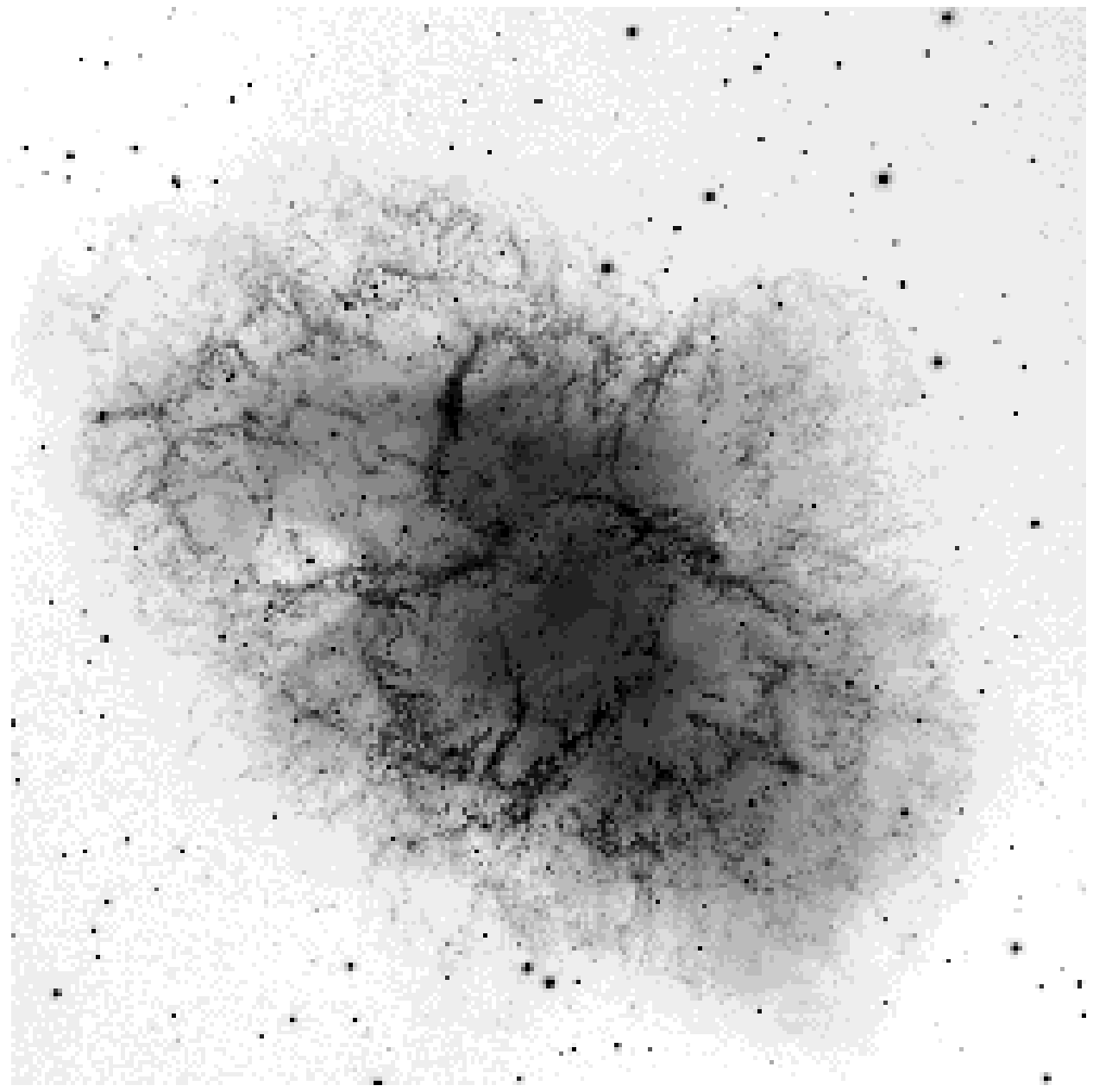} &
		  \includegraphics[scale=0.3]{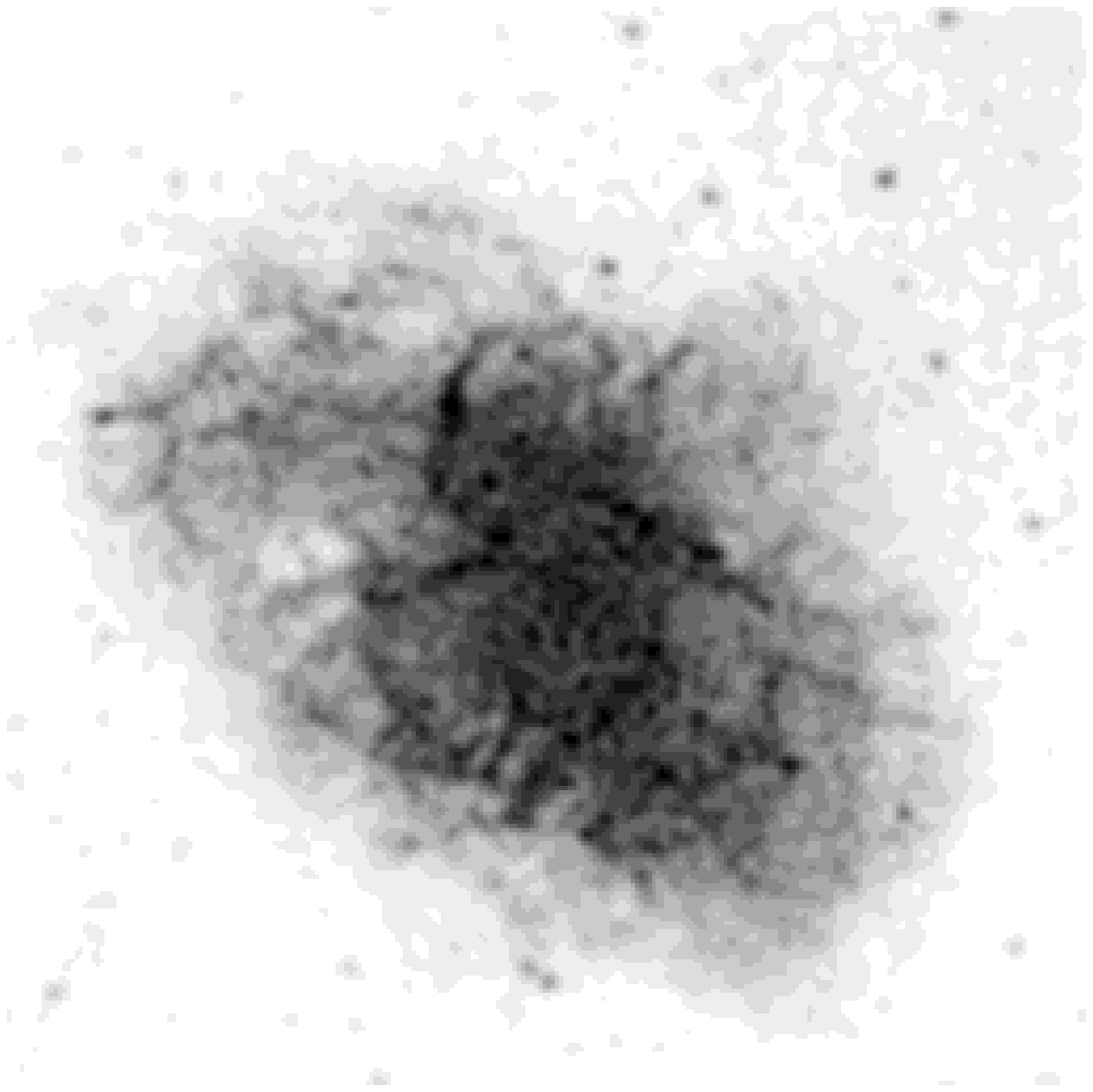}
		\end{tabular}
	\caption{The three objects, represented with reverse gray scale
         (left panels; from up to down nebula, galaxy and Crab), and the
         reconstructions with minimum relative r.m.s. error ($m=15$; right panels).}
	\label{fig1}
\end{figure}

The first experiments are based on $256 \times 256$ HST images of the
planetary nebula NGC7027, the galaxy NGC6946 and the Crab nebula NGC19521. We use
three different integrated magnitudes ($m$) of 10, 12, and 15, not corresponding
to the effective magnitudes of these objects but introduced for
obtaining simulated images with different noise levels. In Fig. \ref{fig1}
we show the three objects in the left panels. In the following, they are
denoted nebula, galaxy and Crab.

\begin{figure}
	\centering
		\begin{tabular}{cc}
			\includegraphics[width=0.18\textwidth]{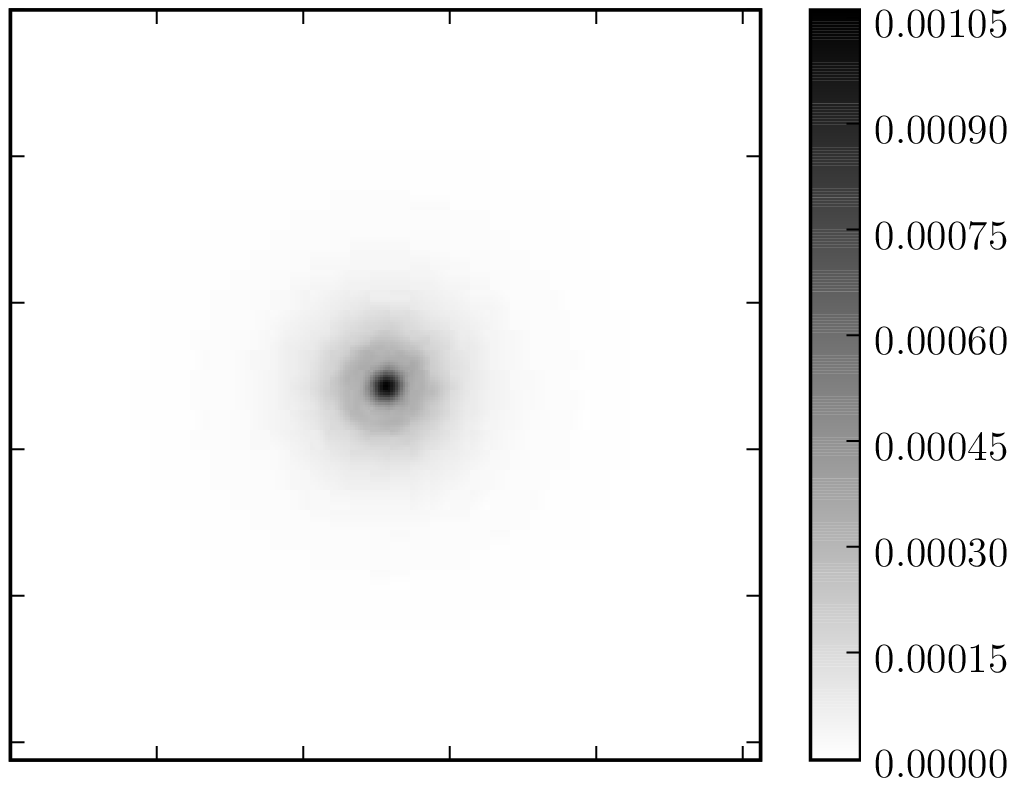} &
		  \includegraphics[width=0.18\textwidth]{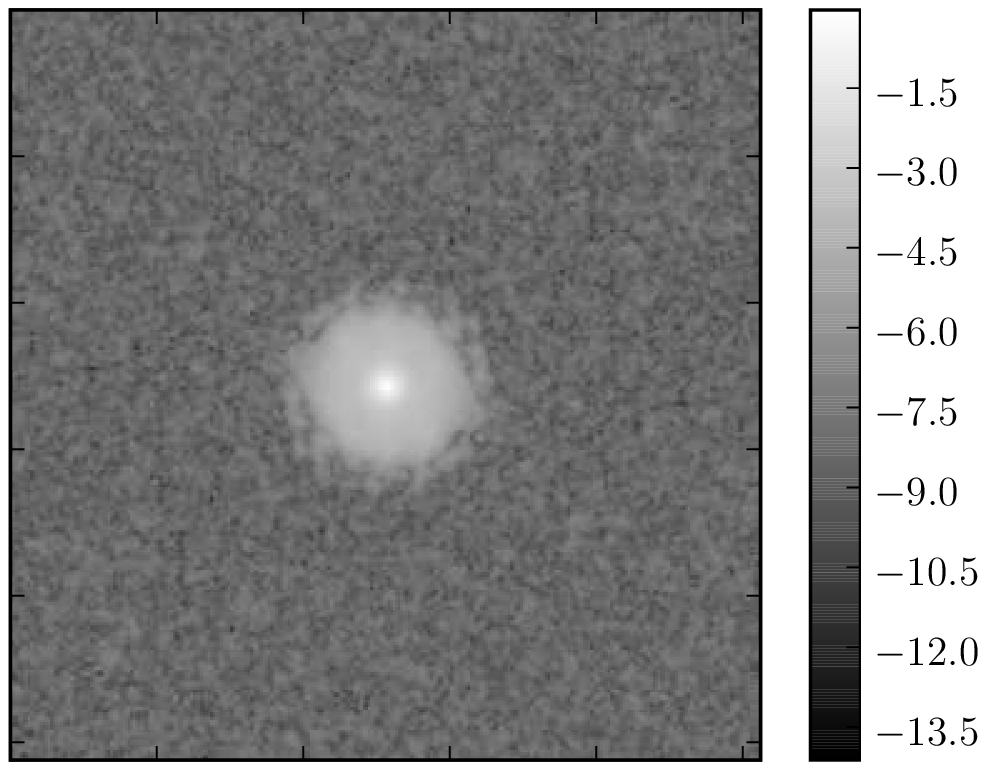}
		\end{tabular}
	\caption{The PSF used in the experiments of single image deconvolution
        (left panel), represented with reverse gray scale, and the
        corresponding MTF (right panel).}
	\label{fig2}
\end{figure}
These objects are convolved with an AO-corrected PSF\footnote{downloaded from
http://www.mathcs.emory.edu/$\sim$nagy/ RestoreTools/index.html} shown
in Fig. \ref{fig2} without zoom, and frequently used in numerical
experiments. The parameters of this PSF (pixel size, diameter of the telescope, etc.)
are not provided. However, it has approximately the same width as
the ideal PSF used in the third experiment reported below and simulated assuming a
telescope of 8.25 m, a wavelength of 2.2 $\mu$m, and a pixel size of 5 mas.

A background of about 13.5 mag arcsec$^{-2}$,
corresponding to observations in K-band, is added to the blurred images and
the results are perturbed with Poisson noise and additive Gaussian noise
with $\sigma=10~e^-/$px. According to the approach
proposed in Snyder et al. (\cite{snyder1}), compensation for readout
noise is obtained in the deconvolution algorithms by adding the constant
$\sigma^2=100$ to the images and the background. In Table \ref{t1},
the performances of RL and SGP are reported, in terms of
iteration numbers needed to obtain the minimum relative r.m.s. error, CPU times, and
speedups provided by the two GPU versions with respect to the serial ones.
The reconstructions corresponding to the minimum relative r.m.s. error, in the case
$m=15$, are shown in the right panels of Fig. \ref{fig1}.

\begin{table*}
\caption{Reconstruction of nebula, galaxy, and Crab as a mosaic of the
reconstructions of four subimages with boundary effect correction. The
number of iterations is the one required for reconstructing each
subdomain, while the reported computational time is the total time required
for the 4 reconstructions.}
\label{t2}
\centering
\begin{tabular}{c c c c c c c c c c c c c}
\hline\hline
          & \multicolumn{4}{c}{Nebula ($m=10$)}  & \multicolumn{4}{c}{Galaxy ($m=10$)}   & \multicolumn{4}{c}{Crab ($m=10$)} \\
Algorithm & It  & Err   & Sec   & SpUp       & It       & Err   & Sec   & SpUp   & It   & Err   & Sec   & SpUp   \\
\hline
RL 			  & 818 & 0.021 & 243.8 &    -        & 10000$^*$ & 0.144 &  2813 &    -   & 4070 & 0.129 &  1146 &    -    \\
RL\_CUDA  & 818 & 0.021 & 12.16 & 20.0        & 10000$^*$ & 0.144 & 141.5 & 19.9   & 4070 & 0.129 & 61.55 & 18.6    \\
SGP 		  &  96 & 0.022 & 35.16 &    -        &       435 & 0.144 & 171.6 &    -   &  129 & 0.129 & 46.42 &    -    \\
SGP\_CUDA &  96 & 0.022 & 3.406 & 10.3        &       435 & 0.148 & 14.41 & 11.9   &  129 & 0.133 & 4.342 & 10.7    \\
\hline
          & \multicolumn{4}{c}{Nebula ($m=12$)}  & \multicolumn{4}{c}{Galaxy ($m=12$)}   & \multicolumn{4}{c}{Crab ($m=12$)} \\
Algorithm & It  & Err   & Sec   & SpUp       & It        & Err   & Sec   & SpUp  & It   & Err   & Sec   & SpUp \\
\hline
RL 			  & 127 & 0.026 & 38.42 &    -        &      2347 & 0.160 & 696.9 &    -   &  696 & 0.137 & 196.5 &    -    \\
RL\_CUDA  & 127 & 0.026 & 2.108 & 18.2        &      2347 & 0.160 & 35.13 & 19.8   &  696 & 0.137 & 10.99 & 17.9    \\
SGP 		  &  21 & 0.026 & 9.563 &    -        &       126 & 0.161 & 51.11 &    -   &   53 & 0.137 & 19.41 &    -    \\
SGP\_CUDA &  21 & 0.026 & 0.874 & 10.9        &       126 & 0.161 & 4.438 & 11.5   &   53 & 0.137 & 1.922 & 10.1    \\
\hline
          & \multicolumn{4}{c}{Nebula ($m=15$)}  & \multicolumn{4}{c}{Galaxy ($m=15$)}   & \multicolumn{4}{c}{Crab ($m=15$)} \\
Algorithm & It  & Err   & Sec   & SpUp       & It        & Err   & Sec   & SpUp  & It   & Err   & Sec   & SpUp   \\
\hline
RL 			  &  96 & 0.064 & 27.58 &    -        &       297 & 0.234 & 89.22 &    -   &   99 & 0.172 & 28.08 &    -    \\
RL\_CUDA  &  96 & 0.064 & 1.703 & 16.2        &       297 & 0.234 & 4.547 & 19.6   &   99 & 0.172 & 1.704 & 16.5    \\
SGP 		  &  10 & 0.061 & 4.234 &    -        &        17 & 0.236 & 7.375 &    -   &    9 & 0.172 & 3.859 &    -    \\
SGP\_CUDA &  10 & 0.061 & 0.407 & 10.4        &        17 & 0.236 & 0.657 & 11.2   &    9 & 0.172 & 0.360 & 10.7    \\
\hline
\end{tabular}
\end{table*}

\begin{figure}
	\centering
		\begin{tabular}{cc}
			\includegraphics[scale=0.3]{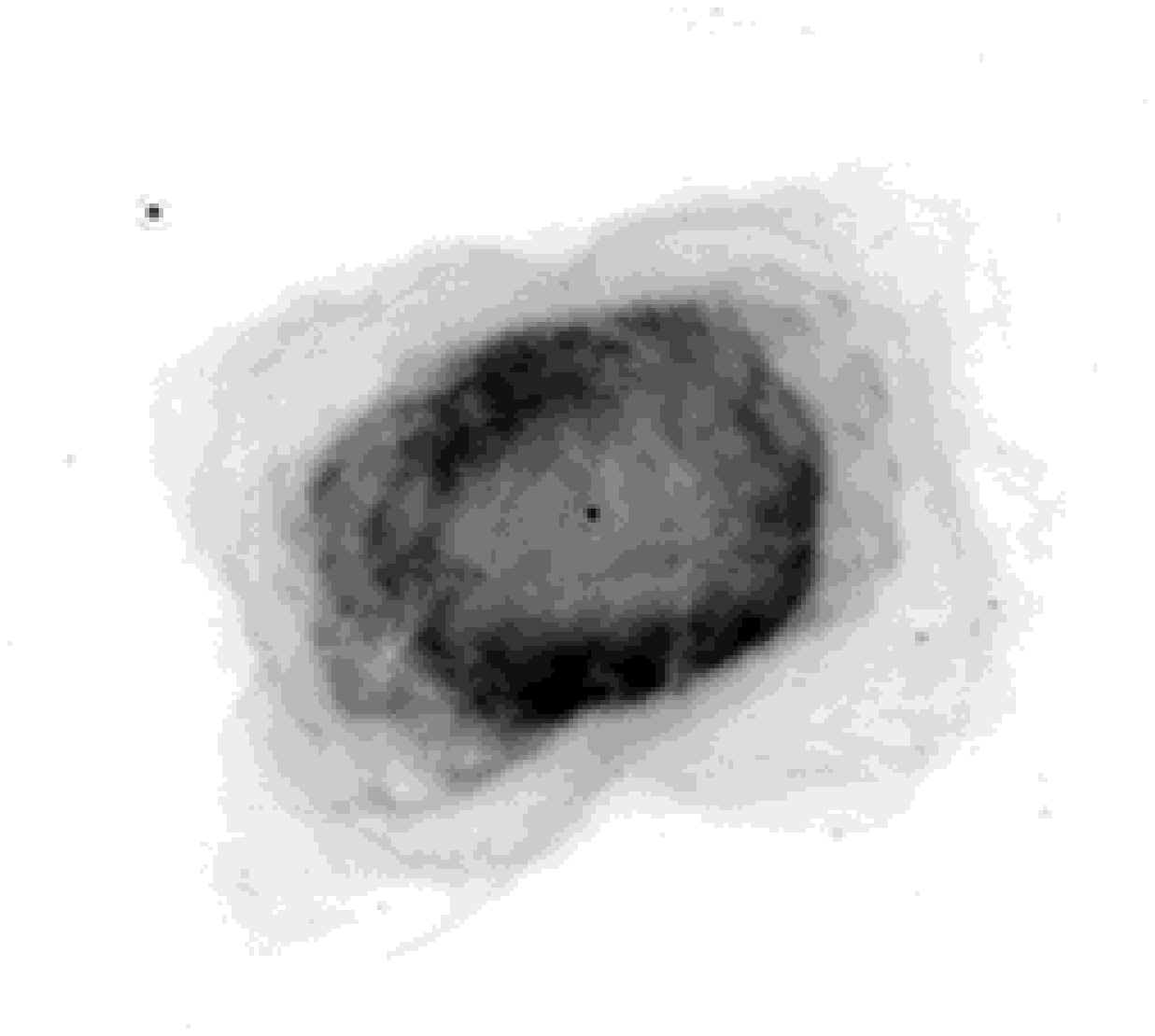} &
		  \includegraphics[scale=0.3]{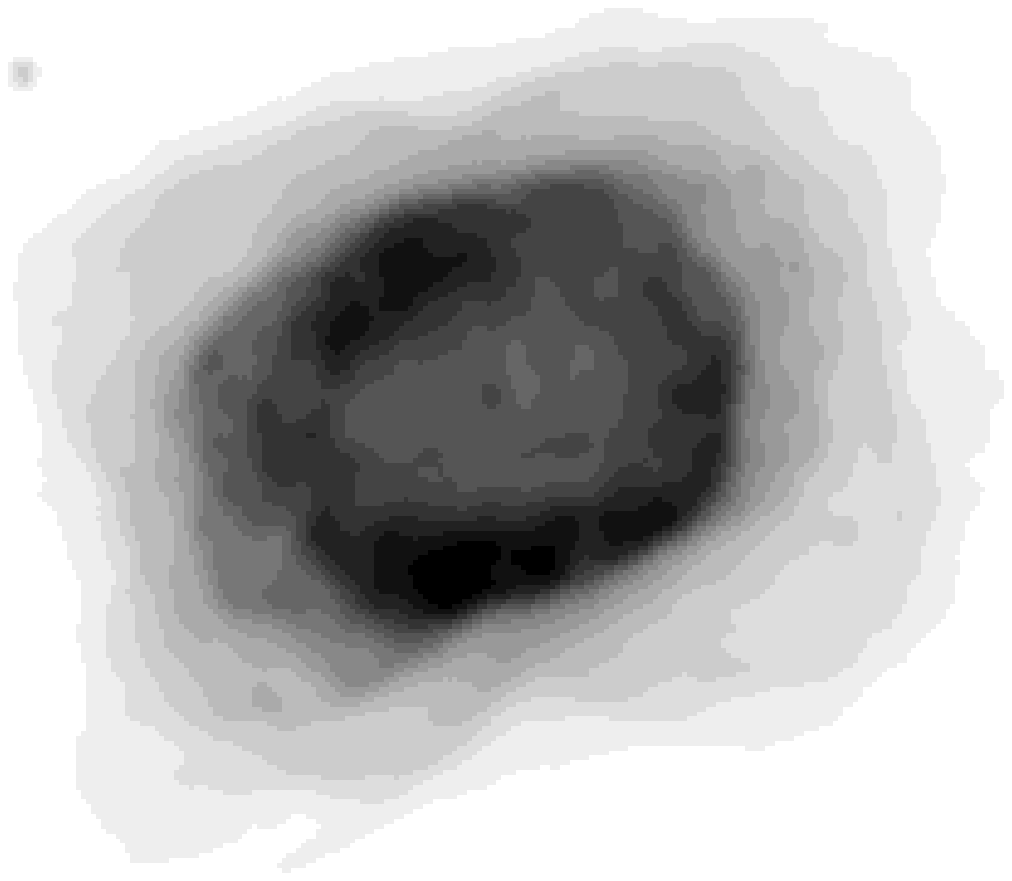} \\
		  \includegraphics[scale=0.3]{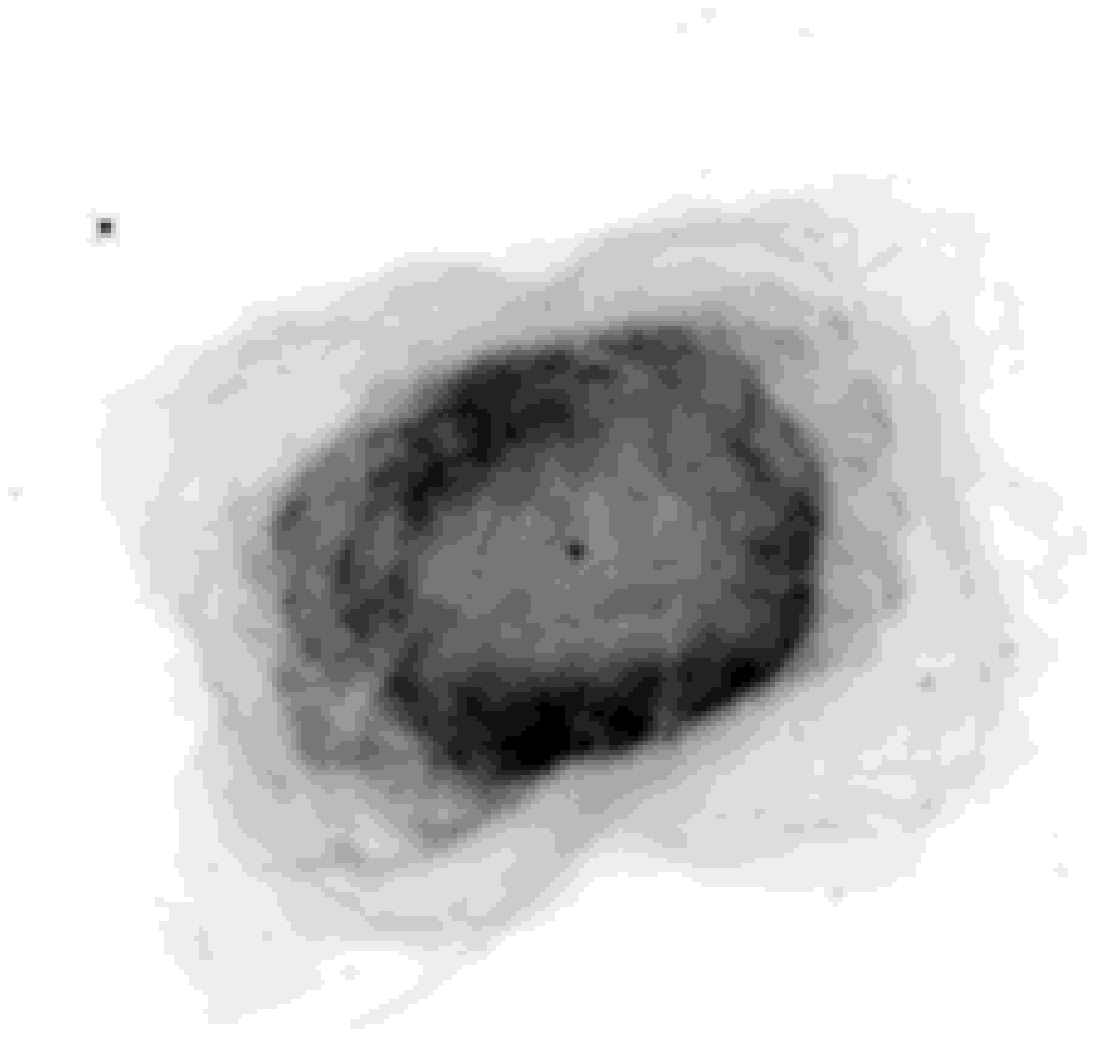} &
		  \includegraphics[scale=0.3]{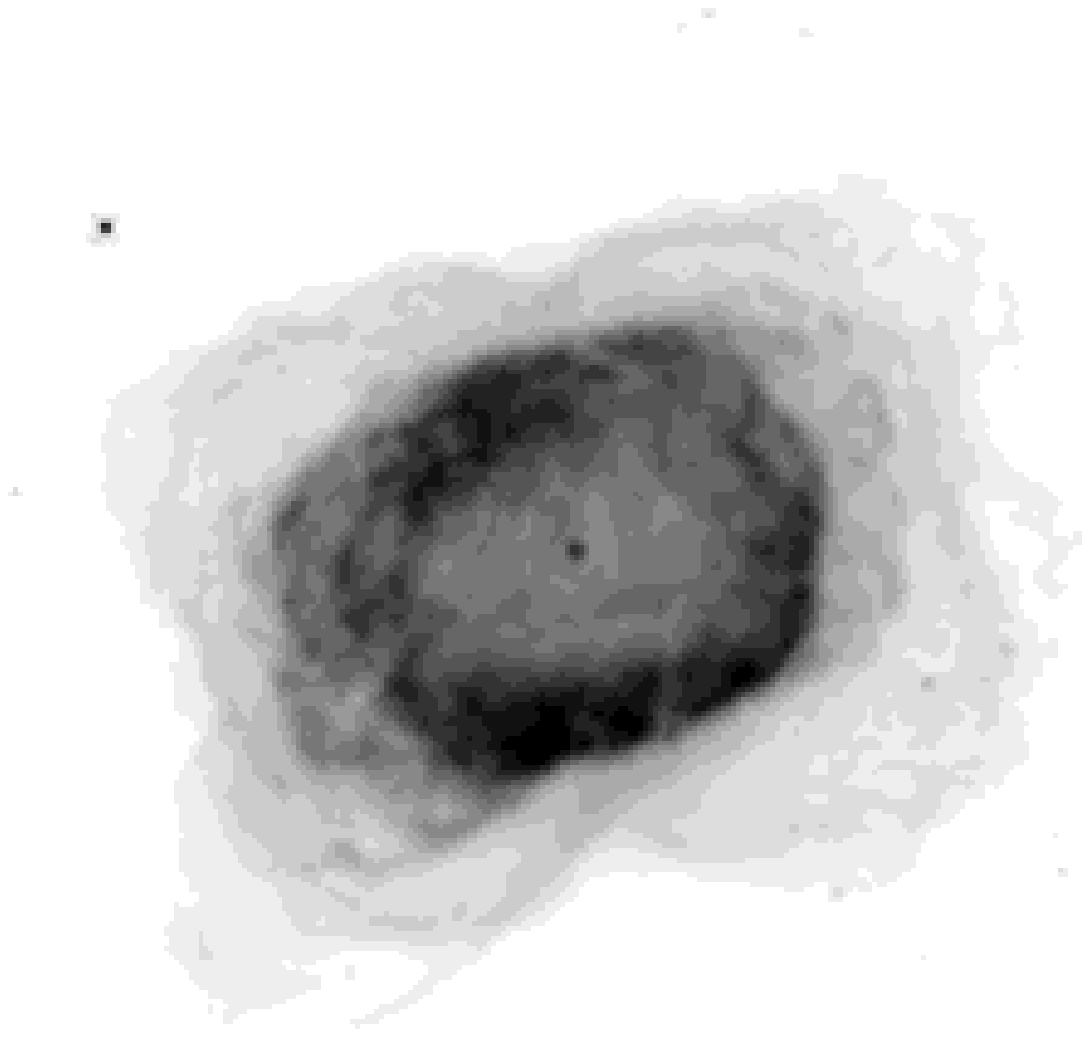}
		\end{tabular}
	\caption{Upper-left panel: the original nebula; upper-right panel: its
        blurred and noisy image in the case $m=10$; lower left panel:
        reconstruction of the global image; lower-left panel:
        reconstruction as a mosaic of four reconstructions of partially
        overlapping subdomains, using the algorithms with boundary
        effect correction.}
	\label{fig3}
\end{figure}

In the second experiment, we use the same datasets created in the previous one
to test the effectiveness of the procedure described in Sect. \ref{bound}
for the reduction of boundary effects. To this aim, the $256 \times 256$
blurred and noisy images are partitioned into four partially overlapping
$160 \times 160$ subdomains. Each one of the four partial images is merged,
by zero-padding, in a $256 \times 256$ array that is used, together with
the original $256 \times 256$ PSF, for the reconstruction of the four parts
of the object by means of the RL and SGP algorithms with boundary effect
correction. From the four reconstructions $128 \times 128$, nonoverlapping
images are extracted and the complete reconstructed image is formed as a
mosaic of them. An example of the result is shown in Fig. \ref{fig3}. By
comparing with the reconstruction of the full image, it is clear that the
mosaic of the four reconstructions does not exhibit visible boundary effects.

The results of this experiment for the three objects are reported in
Table \ref{t2}. The reconstruction error is the relative r.m.s. error
between the mosaic and the original object. By comparing with the results
of Table \ref{t1}, we find that the procedure does not significantly increase
the reconstruction error. We also point out that we choose
the number of iterations corresponding to the {\em global} minimum, i.e.
that providing the best performance on the mosaic of the four reconstructions
obtained in the four subdomains. The computational time is the total time
of the four reconstructions.

The third experiment intends to investigate the speedups achievable by SGP
when varying the size of the images. We adopt the same procedure
used in Ruggiero et al. (2010). The original $256 \times 256$ objects
are convolved with an ideal PSF (described in the second paragraph 
of Sect. \ref{s41}) and perturbed with background and Poisson
noise. Next, images with a larger size are obtained by a Fourier-based
re-binning, i.e. the FFT of the original image is expanded by
zero padding  to a double-sized array and the zero frequency component is
multiplied by four. In this way, the background and the noise level are
approximately unchanged and no new content is introduced at high frequencies.
In particular, no out-of-band noise is introduced and therefore the number of iterations
needed to converge to the best solution is probably underestimated, since we use, 
for any size, the number derived in the case $256 \times 256$.
In this experiment, we consider only the nebula and
the galaxy with two magnitudes 10 and 15. The original images are expanded
up to a size of $2048 \times 2048$. The results are reported in Tables \ref{t3}
and \ref{t4}, where we highlighted both the speedup observed between GPU
and serial implementations (labeled ``Par'') and the one provided by the
use of SGP instead of RL (labeled ``Alg''). We note that the
computational gain achieved by the parallel architecture increases in proportion
to the size of the image. As far as the speedup of SGP with respect to RL is concerned,
strong problem-dependent differences in the number of iterations required
to reach the minimum errors do not lead to a similarly regular behavior.

\begin{table}
\caption{Reconstruction of the nebula NGC7027 with different image sizes.}
\label{t3}
\centering
\begin{tabular}{c c c c c c}
\hline\hline
\multicolumn{6}{c}{$m=10$} \\
\multirow{2}{*}{Algorithm} & \multirow{2}{*}{Size} & \multirow{2}{*}{Err} & \multirow{2}{*}{Sec}  & SpUp   & SpUp  \\
													 &											 &											&												& (Par)  & (Alg) \\
\hline
                  & $256^2$  & 0.051 & 783.9 &       - &       - \\
RL								& $512^2$  & 0.051 &  4527 &       - &       - \\
It = $10000^*$    & $1024^2$ & 0.051 & 17610 &       - &       - \\
									& $2048^2$ & 0.051 & 80026 &       - &       - \\
\hline
                  & $256^2$  & 0.051 & 35.63 &    22.0 &       - \\
RL\_CUDA					& $512^2$  & 0.051 & 69.77 &    64.9 &       - \\
It = $10000^*$    & $1024^2$ & 0.051 & 149.5 &	   118 &       - \\
									& $2048^2$ & 0.051 & 469.1 &	   171 &       - \\
\hline
                  & $256^2$  & 0.052 & 26.14 &       - &    30.0 \\
SGP								& $512^2$  & 0.051 & 143.6 &       - &    31.5 \\
It = 272          & $1024^2$ & 0.051 & 554.0 &       - &    31.8 \\
									& $2048^2$ & 0.051 &  2493 &       - &    32.1 \\
\hline
                  & $256^2$  & 0.052 & 1.797 &    14.5 &    19.8 \\
SGP\_CUDA					& $512^2$  & 0.052 & 3.469 &    41.4 &    20.1 \\
It = 272          & $1024^2$ & 0.052 & 8.016 &    69.1 &    18.7 \\
									& $2048^2$ & 0.052 & 25.66 &    97.2 &    18.3 \\
\hline
\multicolumn{6}{c}{$m=15$} \\
\multirow{2}{*}{Algorithm} & \multirow{2}{*}{Size} & \multirow{2}{*}{Err} & \multirow{2}{*}{Sec}  & SpUp   & SpUp  \\
													 &											 &											&												& (Par)  & (Alg) \\
\hline
                  & $256^2$  & 0.068 & 48.27 &       - &       - \\
RL								& $512^2$  & 0.064 & 278.7 &       - &       - \\
It = 612          & $1024^2$ & 0.062 &  1068 &       - &       - \\
									& $2048^2$ & 0.062 &  4897 &       - &       - \\
\hline
                  & $256^2$  & 0.068 & 2.219 &    21.8 &       - \\
RL\_CUDA					& $512^2$  & 0.064 & 4.109 &    67.8 &       - \\
It = 612          & $1024^2$ & 0.062 & 9.250 &	   115 &       - \\
									& $2048^2$ & 0.062 & 29.13 &	   168 &       - \\
\hline
                  & $256^2$  & 0.068 & 3.016 &		   - &    16.0 \\
SGP								& $512^2$  & 0.064 & 16.95 &		   - &    16.4 \\
It = 31           & $1024^2$ & 0.062 & 65.22 &		   - &    16.4 \\
									& $2048^2$ & 0.061 & 290.8 &		   - &    16.8 \\
\hline
                  & $256^2$  & 0.068 & 0.218 &    13.8 &    10.2 \\
SGP\_CUDA					& $512^2$  & 0.064 & 0.421 &    40.3 &    9.76 \\
It = 31           & $1024^2$ & 0.062 & 1.063 &    61.4 &    8.70 \\
									& $2048^2$ & 0.061 & 3.406 &    85.4 &    8.55 \\
\hline
\end{tabular}
\end{table}

\begin{table}
\caption{Reconstruction of the galaxy NGC6946 with different image sizes.}
\label{t4}
\centering
\begin{tabular}{c c c c c c}
\hline\hline
\multicolumn{6}{c}{$m=10$} \\
\multirow{2}{*}{Algorithm} & \multirow{2}{*}{Size} & \multirow{2}{*}{Err} & \multirow{2}{*}{Sec}  & SpUp   & SpUp  \\
													 &											 &											&												& (Par)  & (Alg) \\
\hline
                  & $256^2$  & 0.293 & 786.0 &       - &       - \\
RL								& $512^2$  & 0.293 &  4545 &       - &       - \\
It = $10000^*$    & $1024^2$ & 0.293 & 17402 &       - &       - \\
									& $2048^2$ & 0.293 & 80022 &       - &       - \\
\hline
                  & $256^2$  & 0.293 & 36.64 &    21.5 &       - \\
RL\_CUDA					& $512^2$  & 0.293 & 67.94 &    66.9 &       - \\
It = $10000^*$    & $1024^2$ & 0.293 & 146.7 &     119 &       - \\
									& $2048^2$ & 0.293 & 463.9 &	   172 &       - \\
\hline
                  & $256^2$  & 0.292 & 88.72 &		   - &    8.86 \\
SGP								& $512^2$  & 0.291 & 484.3 &		   - &    9.38 \\
It = 928          & $1024^2$ & 0.291 &  1854 &		   - &    9.19 \\
			  					& $2048^2$ & 0.291 &  8386 &		   - &    9.54 \\
\hline
                  & $256^2$  & 0.293 & 7.219 &    12.3 &    5.08 \\
SGP\_CUDA					& $512^2$  & 0.293 & 11.14 &    43.5 &    6.10 \\
It = 928          & $1024^2$ & 0.293 & 25.86 &    71.7 &    5.67 \\
									& $2048^2$ & 0.293 & 81.02 &     104 &    5.73 \\
\hline
\multicolumn{6}{c}{$m=15$} \\
\multirow{2}{*}{Algorithm} & \multirow{2}{*}{Size} & \multirow{2}{*}{Err} & \multirow{2}{*}{Sec}  & SpUp   & SpUp  \\
													 &											 &											&												& (Par)  & (Alg) \\
\hline
                  & $256^2$  & 0.311 & 114.9 &       - &       - \\
RL								& $512^2$  & 0.307 & 644.3 &       - &       - \\
It = 1461         & $1024^2$ & 0.306 &  2574 &       - &       - \\
									& $2048^2$ & 0.306 & 11689 &       - &       - \\
\hline
                  & $256^2$  & 0.311 & 5.375 &    21.4 &       - \\
RL\_CUDA					& $512^2$  & 0.307 & 9.656 &    66.7 &       - \\
It = 1461         & $1024^2$ & 0.306 & 22.41 & 	   115 &       - \\
									& $2048^2$ & 0.306 & 68.44 &	   171 &       - \\
\hline
                  & $256^2$  & 0.311 & 3.672 &		   - &    31.3 \\
SGP								& $512^2$  & 0.308 & 20.36 &		   - &    31.6 \\
It = 38           & $1024^2$ & 0.307 & 78.20 &		   - &    32.9 \\
									& $2048^2$ & 0.306 & 354.0 &		   - &    33.0 \\
\hline
                  & $256^2$  & 0.311 & 0.266 &    13.8 &    20.2 \\
SGP\_CUDA					& $512^2$  & 0.307 & 0.531 &    38.3 &    18.2 \\
It = 38           & $1024^2$ & 0.307 & 1.344 &    58.2 &    16.7 \\
									& $2048^2$ & 0.306 & 4.188 &    84.5 &    16.3 \\
\hline
\end{tabular}
\end{table}

\subsection{Multiple images}

We test the efficiency of three algorithms for multiple
image deconvolution, i.e. multiple RL, OSEM, and SGP (applied to multiple
RL), by means of simulated images of the Fizeau interferometer LINC-NIRVANA
(LN, for short; T. Herbst et al., \cite{herbst}) of the Large Binocular
Telescope (LBT). The LN is in an advanced realization phase by a consortium
of German and Italian institutions, led by the Max Planck Institute
for Astronomy in Heidelberg. It is a true imager with a maximum
baseline of 22.8m, thus producing images with anisotropic resolution:
that of a 22.8m telescope in the direction of the baseline and that of a 8.4m
(the diameter of LBT mirrors) in the orthogonal direction. By acquiring
images with different orientations of the baseline and applying suitable
deconvolution methods, it is possible, in principle, to achieve the
resolution of a 22.8m telescope in all directions. The LN will be equipped
with a detector consisting of $2048 \times 2048$ pixels with a pixel
size of about 5mas, corresponding to a F0V of $10" \times 10"$ for
each orientation of the baseline. Since in K-band the resolution of a
22.8m mirror is about 20mas, the detector provides an oversampling of
a factor four.

\begin{figure}
	\centering
		\begin{tabular}{cc}
			\includegraphics[width=0.2\textwidth]{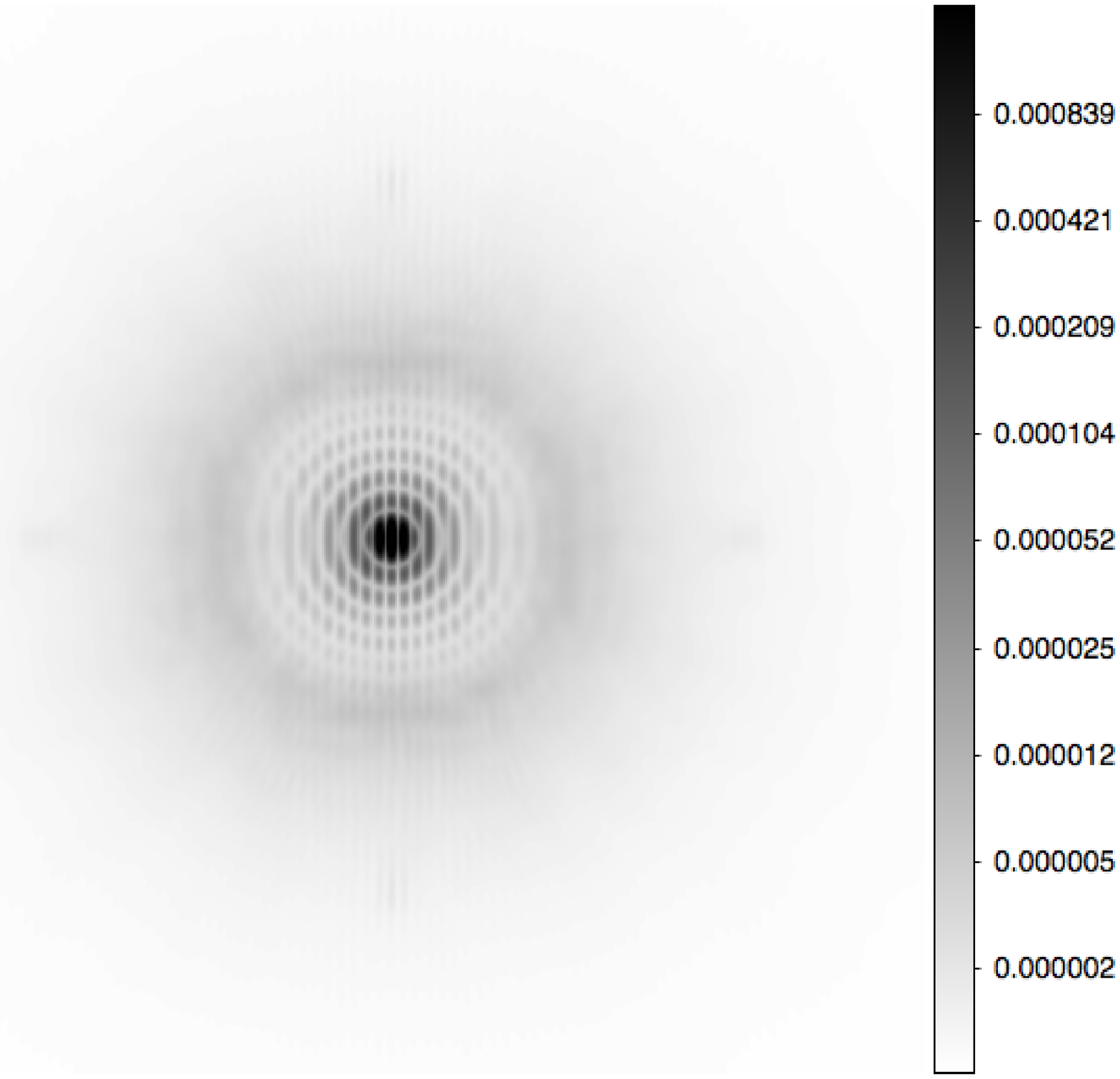} &
		  \includegraphics[width=0.2\textwidth]{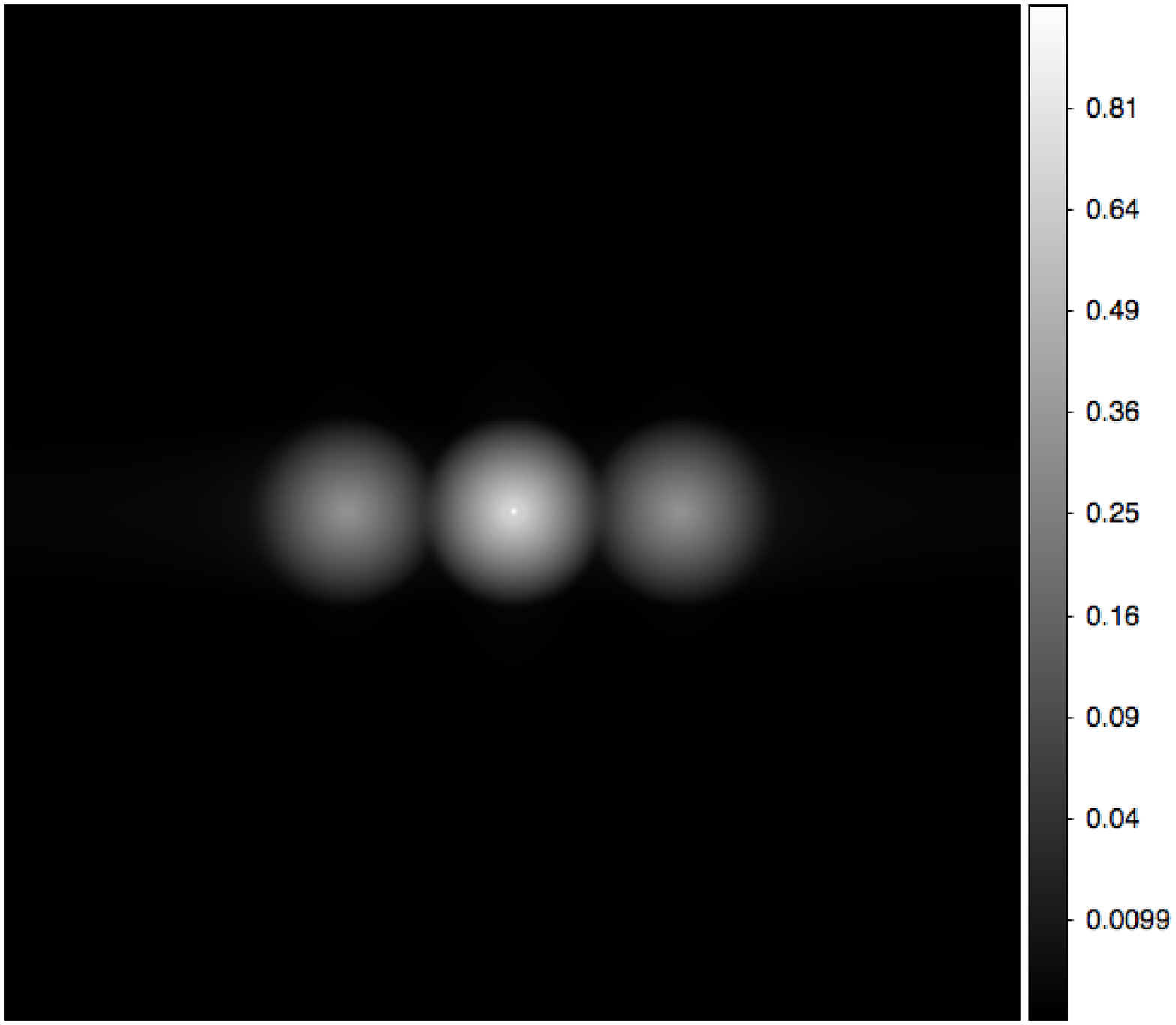}
		\end{tabular}
	\caption{Simulated PSF of LINC-NIRVANA with SR = 70\% (left panel) and
the corresponding MTF (right panel). The PSF is monochromatic in K-band
and is the PSF of a 8.4m mirror (the diameter of the two mirrors of LBT)
modulated by the interferometric fringes. Accordingly, in the MTF the
central disk corresponds to the band of a 8.4m mirror while the two
side disks are replicas due to interferometry.}
	\label{fig4}
\end{figure}

\begin{figure}
	\centering
		\begin{tabular}{cc}
			\includegraphics[width=0.2\textwidth]{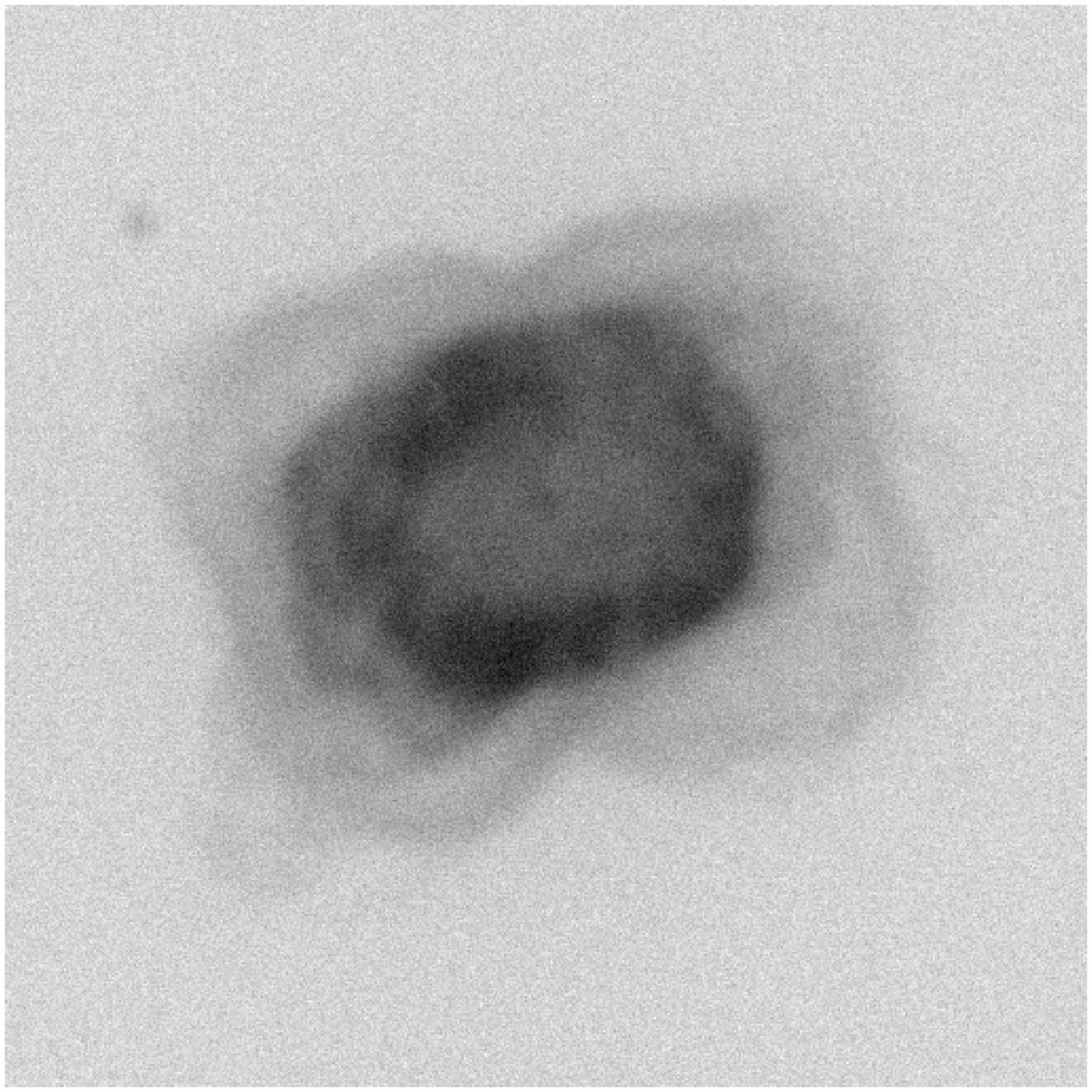} &
		  \includegraphics[width=0.2\textwidth]{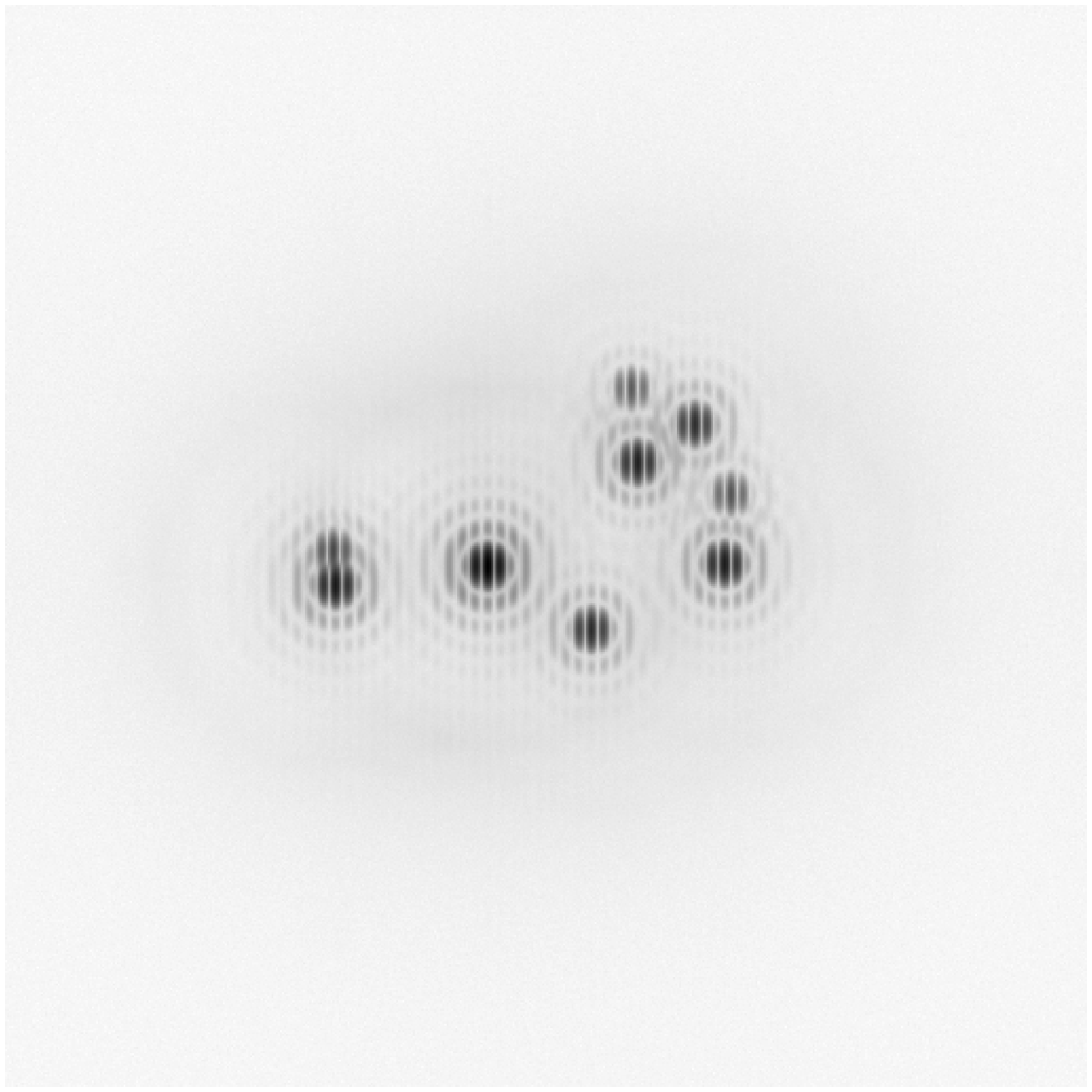}
		\end{tabular}
\caption{Interferometric images (horizontal baseline) of the
$512 \times 512$ Nebula with $m=15$ (left panel) and of the star cluster
(right panel).}
	\label{fig5}
\end{figure}

In our simulations, we use PSFs generated with the code LOST (Arcidiacono
et al. \cite{arcidiacono}); one of them, with SR = 70\% and horizontal
baseline, is shown in Fig. 4 together with the corresponding MTF.
Moreover, we consider two test objects: one is again the nebula NGC7027,
with two magnitudes, 10 and 15, and size $512 \times 512$ (therefore, the
images are noisier than those of the $256 \times 256$ version with the
same integrated magnitude); the other is a model of an open star cluster
based on an image of the Pleiades (star cluster, for short),
consisting of 9 stars with magnitudes ranging from 12.86 to 15.64.
These objects are convolved with three PSFs corresponding to three
equispaced orientations of the baseline, $0^\circ,~60^\circ$, and
$120^\circ$. In the {\it u,v} plane they provide a satisfactory coverage
of the band of a 22.8m telescope (see, for instance, Bertero et al.
\cite{bertero2011}, a review paper where the generation of the images
used in this paper is described in greater detail). The results are
perturbed with a background of about 13.5 mag arcsec$^{-2}$, corresponding
to observations in K-band, and with both Poisson and Gaussian noises
($\sigma=10~e^-/$px). In Fig. 5, we show one interferometric image of the
nebula, with magnitude 15, and one interferometric image of
the star cluster, both with a horizontal baseline.

\subsubsection{Diffuse objects}

\begin{table}
\caption{Reconstruction of the nebula using three equispaced $512\times512$
images.}
\label{t5}
\centering
\begin{tabular}{c c c c c}
\hline\hline
\multicolumn{5}{c}{$m=10$} \\
Algorithm  & It   & Err   & Sec   & SpUp \\
\hline
RL 			   & 3401 & 0.032 &  4364 &    -   \\
RL\_CUDA 	 & 3401 & 0.032 & 48.00 & 90.9   \\
OSEM  	   & 1133 & 0.032 &  1602 &    -   \\
OSEM\_CUDA & 1133 & 0.032 & 18.59 & 86.2   \\
SGP 		   &  144 & 0.033 & 220.7 &    -   \\
SGP\_CUDA  &  144 & 0.033 & 3.563 & 61.9   \\
\hline
\multicolumn{5}{c}{$m=15$} \\
Algorithm  & It  & Err   & Sec   & SpUp \\
\hline
RL 			   & 353 & 0.091 & 441.5 &    -  \\
RL\_CUDA 	 & 353 & 0.091 & 4.937 & 89.4  \\
OSEM		   & 117 & 0.091 & 165.7 &    -  \\
OSEM\_CUDA & 117 & 0.091 & 2.062 & 80.4  \\
SGP 		   &  16 & 0.087 & 26.14 &    -  \\
SGP\_CUDA  &  16 & 0.087 & 0.546 & 47.9  \\
\hline
\end{tabular}
\end{table}

We now provide the results obtained in the case of the nebula
with two magnitudes, 10 and 15. The stopping rule is given again by the
minimum r.m.s. error. We first consider deconvolution without
correction for edge artifacts because the object is within the
image domain. The results are reported in Table \ref{t5}. If we compare
the behaviors of single image and multiple
image RL, we find that in the second case a larger number of iterations
is required, owing to the difficulty in combining the resolutions of the
different images to get a unique high-resolution reconstruction. Moreover,
the greater cost per iteration has two causes: the first is that
the size is $256 \times 256$ in the single case and $512 \times 512$
in the multiple image case; the second is that one single image iteration
requires 4 FFTs, while one multiple image iteration, with three images,
requires 10 FFTs.

The results confirm that the speedup provided by OSEM with respect to
multiple RL is about 2.5 with a reduction by a factor 3 in the number of
iterations (see Sect. 2.3), although the speedup provided by SGP with
respect to OSEM of a factor between 6 and 7 is interesting. 
This speedup presumably decreases as the number of images increases, but a speedup
of about 20 is provided by OSEM in the case of 26 images, a number that
presumably will never be achieved in the case of LN. Therefore, one can
conclude that SGP can be recommended for the deconvolution of LN images.
Our CUDA implementations provide an additional speedup of about 80/90
for RL and OSEM, while smaller factors are observed for SGP.

When testing the accuracy of the deconvolution methods with boundary effect
correction, we follow the same procedure used in the single image case, i.e.
the images are partitioned into four partially overlapping subimages, the
methods with boundary effect correction are applied and the final
reconstruction is obtained as a mosaic of the four partial reconstructions.
The results are reported in Table \ref{t6} and confirm the results
obtained in the single image case.

\begin{table}
\caption{Reconstruction of the nebula as a mosaic of four reconstructed
subimages with boundary effect correction, also in the case of three
equispaced images.}
\label{t6}
\centering
\begin{tabular}{c c c c c}
\hline\hline
\multicolumn{5}{c}{$m=10$} \\
Algorithm  & It   & Err   & Sec   & SpUp \\
\hline
RL 			   & 2899 & 0.034 & 13978 &    -  \\
RL\_CUDA 	 & 2899 & 0.034 & 174.2 & 80.2  \\
OSEM		   &  950 & 0.034 &  5447 &    -  \\
OSEM\_CUDA &  950 & 0.034 & 64.03 & 85.1  \\
SGP 		   &  160 & 0.034 & 873.3 &    -  \\
SGP\_CUDA  &  160 & 0.034 & 15.45 & 56.5  \\
\hline
\multicolumn{5}{c}{$m=15$} \\
Algorithm  & It & Err    & Sec   & SpUp \\
\hline
RL 			   & 243 & 0.094 &  1174 &    -  \\
RL\_CUDA 	 & 243 & 0.094 & 15.28 & 76.8  \\
OSEM		   &  81 & 0.094 & 479.1 &    -  \\
OSEM\_CUDA &  81 & 0.094 & 5.939 & 80.7  \\
SGP 		   &  11 & 0.087 & 69.88 &    -  \\
SGP\_CUDA  &  11 & 0.086 & 1.532 & 45.6  \\
\hline
\end{tabular}
\end{table}

\subsubsection{Point-wise objects}

In this case, iterations are pushed to convergence and therefore the stopping
rule is given by the condition in Eq. (\ref{convergence}); we use different values
of {\it tol}, specifically $10^{-3},~10^{-5}$, and $10^{-7}$. In order to
measure the quality of the reconstruction, we introduce an average relative
error of the magnitudes defined by
\begin{equation}
{\rm{av\_rel\_er}} = \frac{1}{q}\sum_{j=1}^{q}\frac{|m_j-{\widetilde m}_j|}
{\widetilde m_j}~~,
\label{averror}
\end{equation}
where $q$ is the number of stars (in our case $q=9$) and ${\widetilde m}_j$
and $m_j$ are respectively the true and the reconstructed magnitudes. The
results are reported in Table \ref{t7}.

We first point out that, as in the previous cases, we constrain the parallel
codes to perform the same number of iterations as the serial ones. This
constraint is introduced because the FFT does not have the same precision
in the two cases, as already discussed. As a result, the two
implementations of the same algorithm do not provide the same error for
the same number of iterations. This effect presumably will be removed when
a double-precision FFT becomes available for GPU.

Next, we find, as expected, that the number of iterations increases with
decreasing values of $\it tol$. However, the increase in computation time is
not compensated by a significant decrease in the accuracy of the
reconstructed magnitudes. For ${\it tol}=10^{-3}$, the accuracy of the
estimated magnitudes might already be satisfactory. We observe, however,
that with this milder tolerance the accuracy provided by the three algorithms
is not the same. Multiple RL and OSEM seem to be slightly more accurate.
The accuracy of all algorithms is essentially the same for the smaller
tolerances.

\begin{table}
\caption{Reconstruction of the star cluster with three $512\times512$
equispaced images. The error is the average relative error in the magnitudes defined
in Eq. (\ref{averror}).}
\label{t7}
\centering
\begin{tabular}{c c c c c}
\hline\hline
\multicolumn{5}{c}{$tol$ = 1e-3} \\
Algorithm  & It   & Err     & Sec   & SpUp  \\
\hline
RL 			   &  319 & 2.39e-4 & 393.4 &    -  \\
RL\_CUDA 	 &  319 & 2.38e-4 & 4.641 & 84.8  \\
OSEM 		   &  151 & 1.63e-4 & 220.8 &    -  \\
OSEM\_CUDA &  151 & 1.62e-4 & 2.421 & 91.2  \\
SGP 			 &   71 & 1.35e-3 & 97.80 &    -  \\
SGP\_CUDA  &   71 & 1.29e-3 & 1.641 & 59.6  \\
\hline
\multicolumn{5}{c}{$tol$ = 1e-5} \\
Algorithm  & It   & Err     & Sec   & SpUp \\
\hline
RL 			   & 1385 & 6.65e-5 &  1703 &    -  \\
RL\_CUDA 	 & 1385 & 6.64e-5 & 19.38 & 87.9  \\
OSEM 		   &  675 & 5.64e-5 & 980.6 &    -  \\
OSEM\_CUDA &  675 & 5.64e-5 & 10.75 & 91.2  \\
SGP 			 &  337 & 5.89e-4 & 455.2 &    -  \\
SGP\_CUDA  &  337 & 1.79e-4 & 7.187 & 63.3  \\
\hline
\multicolumn{5}{c}{$tol$ = 1e-7} \\
Algorithm  & It   & Err     & Sec   & SpUp  \\
\hline
RL 			   & 7472 & 5.64e-5 &  9180 &    -  \\
RL\_CUDA 	 & 7472 & 5.98e-5 & 104.8 & 87.6  \\
OSEM 		   & 3750 & 6.13e-5 &  5442 &    -  \\
OSEM\_CUDA & 3750 & 5.98e-5 & 59.52 & 91.4  \\
SGP 			 &  572 & 7.37e-5 & 772.6 &    -  \\
SGP\_CUDA  &  572 & 7.05e-5 & 12.20 & 63.3  \\
\hline
\end{tabular}
\end{table}

As a final experiment, we consider the reconstruction of a binary with
high dynamic range (Bertero et al. \cite{bertero2011}). It consists
of a primary with $m_1=10$ (denoted as $S_1$) and a secondary with
$m_2=20$ (denoted as $S_2$). The distance between the two stars is
45mas (i.e. 9 pixels for the LINC-NIRVANA detector) and the axis of the
binary forms an angle of $23^\circ$ with the direction of the
baseline of the first image. Three equispaced images are generated
as in the case of the star cluster, using the same PSFs and the same
background.

In this experiment, we need a very small tolerance, i.e. $tol = 10^{-7}$,
in order to allow SGP to detect the faint secondary. The reason is
presumably that SGP requires a projection onto the non-negative orthant,
and the existence of this projection can make degrade the appearance
of the secondary. In all cases, the results reported in Table \ref{t8} are
interesting and demonstrate that the magnitude of the secondary can also
be estimated with a sufficient accuracy in a reasonable computation
time.

\begin{table}
\caption{Reconstruction of the binary with high dynamic range (image size:
$256\times256$).}
\label{t8}
\centering
\begin{tabular}{c c c c}
\hline\hline
\multicolumn{4}{c}{$tol$ = 1e-7} \\
Algorithm & It & Sec & SpUp \\
\hline
RL 			   & 30765 &  6108 &    -  \\
RL\_CUDA 	 & 30765 & 292.9 & 20.9  \\
OSEM 		   & 14291 &  3216 &    -  \\
OSEM\_CUDA & 14291 & 156.0 & 20.6  \\
SGP 			 &  2073 & 482.8 &    -  \\
SGP\_CUDA  &  2073 & 28.59 & 16.9  \\
\hline
\multicolumn{4}{c}{Magnitude} \\
Algorithm & Star & Real & Reconstructed \\
\hline
\multirow{2}{*}{RL} 			 & S1 & 10 & 10.0001 \\
													 & S2 & 20 & 20.1841 \\
\hline
\multirow{2}{*}{OSEM} 		 & S1 & 10 & 10.0001 \\
													 & S2 & 20 & 20.0919 \\
\hline
\multirow{2}{*}{SGP} 			 & S1 & 10 & 10.0001 \\
													 & S2 & 20 & 20.2683 \\
\hline
\end{tabular}
\end{table}

\section{Discussion}

The codes of the algorithms presented and discussed in this paper can be freely
downloaded\footnote{at the URL http://www.unife.it/prin/software}. A MATLAB code of SGP for single image deconvolution is
also available at the same URL, which enables one to compare the IDL and MATLAB implementations on one's own computer.

The paper is based on the RL algorithm and its
generalizations to boundary effect correction and multiple image
deconvolution being scaled gradient methods, where the scaling is provided
by the current iterate. Therefore, it is possible to attempt to improve the efficiency 
of these algorithms in the framework of the SGP approach
proposed in Bonettini et al. (\cite{bonettini}). As already shown in
that paper, the SGP version of RL provides a considerable increase in
efficiency.

The results given in the previous section demonstrate that SGP allows
a significant speedup of all the RL-type algorithms considered in
this paper, even if the speedup depends considerably on the specific
object to be reconstructed and, for a given object, on the noise level;
it ranges from about 4 in the case of multiple images
of the star cluster (Table \ref{t7}), to more than 30, in
the case of a single image of the galaxy
(Table \ref{t4}). A more accurate investigation of the speedup
achievable would require application to a broader data set of
astronomical objects as well as to images with different noise levels
and noise realizations. In all cases, we believe that the results presented in
this paper are sufficient to demonstrate that SGP is a valuable
acceleration of RL-like algorithms and that in several cases it allows
a considerable reduction in computational time.

The speedup provided by GPU implementation is
consistent with the results reported in Ruggiero et al. (\cite{ruggiero}).
The speedup of RL-algorithms is greater than that of SGP-algorithms
because the main computational kernel of RL is FFT, while SGP is
also based on the computation of steplengths, etc. Nevertheless, the gain with respect
to RL is very significant, in some cases, allowing us to
deconvolve a $2048 \times 2048$ image in a few seconds.
We analyzed the use of a multi-thread FFTW in our C
implementation of the SGP algorithm, obtaining an improvement in the performance
with respect to the corresponding serial code, which is however definitely lower than that achieved
with the CUDA version.

We conclude by emphasizing that in this paper we have considered a
maximum likelihood approach for the image deblurring problems and used an early
stopping of the iterative procedures to mimic a regularization effect.
However, the SGP method can also be applied to
regularized deconvolution in the framework of a Bayesian approach. The
main problem would then be to decide on a rule to determine a suitable scaling for
a given regularization function, since the scaling should depend on
this function. Such a rule can be provided by the split-gradient method
(SGM) proposed by Lant\'eri et al. (\cite{lanteri0,lanteri}), which can
be considered an improvement of the one-step late (OSL) method proposed
by Green (\cite{green}): the OSL scaling does not always yield positive values, 
while the SGM scaling does.

The scaling of SGM combined with SGP has been already tested in the
case of Poisson denoising (Zanella et al. \cite{zanella}) and Poisson deblurring (Staglian\`o et al. \cite{stagliano}), which are both based on
edge-preserving regularization. In both cases, this combination
leads to very efficient algorithms. The SGM scalings can also be designed
for other kinds of regularization, and for a discussion of the case for
Poisson data we refer to Bertero et al. (\cite{bertero2009,bertero2011}).

Work is in progress to develop a library of SGP
algorithms for Poisson data deconvolution with a number of different
kinds of regularization.

\begin{acknowledgements}
This work has been partially supported by MIUR (Italian Ministry for University and
Research), PRIN2008 ``Optimization Methods and Software for Inverse Problems'', grant 2008T5KA4L, 
and by INAF (National Institute for Astrophysics) under the contract TECNO-INAF 2010 
``Exploiting the adaptive power: a dedicated free software to optimize and maximize the 
scientific output of images from present and future adaptive optics facilities''.
\end{acknowledgements}

\end{document}